\documentclass[12pt]{article}
\usepackage{setspace}
\usepackage{amsmath,amssymb}
\usepackage{color}
\usepackage{hyperref}
\usepackage{array}
\usepackage{booktabs}
\numberwithin{equation}{section}
\usepackage{graphicx}

\usepackage{soul}

\newcommand{\beq}{\begin{equation}}
\newcommand{\eeq}[1]{\label{#1}\end{equation}}
\newcommand{\bea}{\begin{eqnarray}}
\newcommand{\eea}[1]{\label{#1}\end{eqnarray}}
\renewcommand{\Im}{{\rm Im}\,}
\renewcommand{\Re}{{\rm Re}\,}

\def\mL{\mathcal{L}}

\def\mM{\mathcal{M}}
\def\mO{\mathcal{O}}

\def\mM{\mathcal{M}}

\def\mI{\mathcal{I}}
\def\mC{\mathcal{C}}

\def\zb{\bar{z}}
\def\wb{\bar{w}}

\def\qb{\bar{q}}

\def\pa{\partial}

\def\g5{\gamma_5}

\def\ut{\tilde{u}}
\def\vt{\tilde{v}}

\def\b[#1]{\bold{#1}}
\def\bb[#1]{\overline{\bold{#1}}}
\def\bs[#1,#2]{\bold{#1}_{#2}}
\def\bbs[#1,#2]{\overline{\bold{#1}}_{#2}}

\def\s2{\sigma_2}

\def\gammaflat{ \gamma_{z\zb}}
\def\gammaflatt{ \gamma^{z\zb}}

\def\Tsoft{T_{\text{soft}}}
\def\Thard{T_{\text{hard}}}

\def\omegak{\omega_k}

\def\paz{\pa_z}

\def\pazb{\pa_{\zb}}

\def\ketd[#1]{\ket{#1}_{\text{dressed}}}
\def\brad[#1]{\bra{#1}_{\text{dressed}}}
\def\ketas[#1]{\ket{#1}_{\text{Asymptotic}}}
\def\braas[#1]{\bra{#1}_{\text{Asymptotic}}}

\def\Nt{\tilde{N}}
\def\ImNt{\Im \Nt }
\def\ReNt{\Re \Nt }

\def\ImNtz{\Im \Nt (z,\zb)\,}
\def\ReNtz{\Re \Nt (z,\zb)\, }

\def\eval{\, \! \big|}

\begin{document}
	\setlength{\topmargin}{-1cm} 
	\setlength{\oddsidemargin}{-0.25cm}
	\setlength{\evensidemargin}{0cm}

\begin{titlepage}
\hfill CERN-TH-2019-069
\begin{center}

\vskip 4 cm

{\LARGE \bf  Properties of Dual Supertranslation  Charges in \vskip .1in Asymptotically Flat Spacetimes}

\vskip 2 cm

{Uri Kol$^a$  and Massimo Porrati$^{a,b}$}

\vskip .75 cm

{$^a$ \em Center for Cosmology and Particle Physics\\
	Department of Physics, New York University \\
	726 Broadway, New York, NY 10003, USA}
	
	\vspace{12pt}

{$^b$ \em CERN TH,
CH 1211, Geneva 23, Switzerland\footnote{Until August 31, 2019; on sabbatical leave from NYU.}}

\end{center}

\vskip 1.25 cm

\begin{abstract}
\noindent
We study several properties of some new charges of asymptotically flat spacetimes. These  {\em
dual supertranslation charges} are akin to the magnetic large $U(1)$ charges in QED. In this paper we find the 
symmetries associated with these charges and show that the global dual supertranslation charge is topological because it is 
invariant under globally defined, smooth variations of the asymptotic metric. We also exhibit spacetimes where the charge does
 not vanish and we find dynamical processes that interpolate between regions with different values of these charges. 
\end{abstract}
\end{titlepage}
\newpage

\section{Introduction}

New {\em dual gravitational charges} of asymptotically flat spacetimes were introduced in~\cite{Godazgar:2018qpq, Godazgar:2018dvh}.
They  are akin to the magnetic dual of the large $U(1)$ gauge charges~\cite{He:2014cra,Strominger:2013lka,Strominger:2015bla,Strominger:2017zoo} (see also~\cite{shiu})  so we shall call them  ``dual supertranslation charges''  henceforth. The symmetries
they generate are not explicit in the derivation of~\cite{Godazgar:2018qpq, Godazgar:2018dvh}   
so we shall begin our paper by reporting
 on an independent line of research that we have been pursuing recently and that arrives at them in a different
manner, which makes their action on the phase space of the theory more manifest. 
Another important question about these charges concerns the extent to which one can draw the parallel with large 
magnetic $U(1)$ transformations. The global magnetic charge is topological in nature since it does not change under small, globally-defined 
 deformations of the gauge potential, it is quantized in units set by the electric charge and it partitions the space of
 gauge field configurations on the celestial sphere into discrete, topologically distinct sectors. In this paper we
 will provide evidence that the global dual supertranslation charge is also topological.
 
Let us begin with some standard definitions, the first of which is the expansion of an asymptotically flat metrics 
 around future null infinity $\mathcal{I}^+$:
\begin{equation}\label{futureMetric}
\begin{aligned}
ds^2 &= -du^2 -2dudr+2r^2 \gammaflat dz d\zb \\
&+
\frac{2m_B}{r} du^2 +r C_{zz} dz^2 +r C_{\zb\zb} d\zb ^2 -2U_z du dz -2 U_{\zb} du d\zb \\
&+ \frac{1}{r}\left(
\frac{4}{3}\left(N_z+u \paz m_B \right) - \frac{1}{4} \paz \left(C_{zz}C^{zz}\right)
\right) du dz + c.c.
\\
&+ \dots ,
\end{aligned}
\end{equation}
where $u$ is the retarded time, 
\begin{equation}
\gammaflat = \frac{2}{(1+z \zb)^2}
\end{equation}
is the round metric on the unit $S^2$,
\begin{equation}
U_z = -\frac{1}{2} D^z C_{zz}
\end{equation}
and the dots indicate subleading terms in the expansion around $r= \infty$. The symbol $m_B$ denotes the Bondi mass aspect and $N_z$ the angular momentum aspect. The Bondi news, given by
\begin{equation}
N_{zz} = \pa_u C_{zz},
\end{equation}
characterizes gravitational radiation.

In \cite{Strominger:2013jfa,He:2014laa} it was argued that in order to achieve the correct Dirac bracket on the radiative modes at future null infinity $\mI^+$
\begin{equation}\label{DiracBra}
\{ T (f) , C_{zz}  \} =f \pa_u C_{zz} - 2 D_z^2f = \mL_f C_{zz},
\end{equation}
where $T (f)$ is the generator of $\text{BMS}^+$ supertranslations, the following boundary conditions have to be imposed
\begin{equation}\label{StandradBC}
\Big[ D_{\zb}^2 C_{zz}  -  D_z^2 C_{\zb\zb} \Big]_{\mI^+_{-}}=0,
\end{equation}
\begin{equation}\label{gravRadCond}
N_{zz}\eval  _{\mI^+_{-}}=0.
\end{equation}
The condition \eqref{StandradBC} fixes the coefficient of the $D_z^2f$ term in equation \eqref{DiracBra} to be $(-2)$. Without it, this coefficient would be $-1$ instead of $-2$. The reason for this discrepancy lies in the way we count zero-momentum modes. For non-zero momentum the graviton has two degrees of freedom corresponding to the two polarization modes with respect to its momentum. However, at zero momentum the two polarization modes cannot be distinguished so that in fact only one degree of freedom survives. The constraint \eqref{StandradBC} relates the two polarization modes of the zero-momentum graviton and therefore reduces the number of degrees of freedom from two to one, as required. The numerical factor in front of the $D_z^2f$ term in \eqref{DiracBra} precisely accounts for this effect.

In this paper we show that there is a more general class of boundary conditions that is consistent with the correct Dirac brackets,  to wit: 
\begin{equation}\label{NewBC}
i \alpha \Big[
D_{\zb}^2 C_{zz} -D_z^2 C_{\zb\zb}
\Big]_{\mI^+_{-}}
+\beta 
\Big[
D_{\zb}^2 C_{zz} +D_z^2 C_{\zb\zb}
\Big]_{\mI^+_{-}}
=0,
\end{equation}
\begin{equation}
N_{zz}\eval  _{\mI^+_{-}}=0.
\end{equation}
Here $\alpha$ and $\beta$ are real but otherwise unconstrained coefficients. 
The solution to \eqref{NewBC} is given by
\begin{equation}\label{boundaryGraviton}
 C_{zz} \eval _{\mI^+_{-}} = D_z^2 C (z , \zb),
\end{equation}
where $C(z,\zb)$ is a \emph{complex} function, the \emph{boundary graviton}.
To see this we plug \eqref{boundaryGraviton} into \eqref{NewBC} to get the following constraint
\begin{equation}\label{constraint}
-\alpha \, \Im C + \beta \, \Re C =0,
\end{equation}
which relates the real and imaginary parts of the boundary graviton.
For generic values of $\alpha$ and $\beta$ both parts are different than zero.
By setting $\beta$ to zero we recover the standard boundary condition  \eqref{StandradBC} which implies that $C$ is a real function. In the same way we can isolate the imaginary part of $C$ by setting its real part to zero using the alternative boundary condition
\begin{equation}\label{orthogonalBC}
\Big[  D_{\zb}^2 C_{zz}  +  D_z^2 C_{\zb\zb}  \Big]_{\mI^+_{-}} = 0,
\end{equation}
which corresponds to the choice $\alpha=0$.
The nature of the real and imaginary components of the boundary graviton can be understood using their transformation law under the antipodal map, which is equivalent, up to rotations on the two sphere, to
\begin{equation}
z \longleftrightarrow \zb.
\end{equation}
Under the antipodal map the two components transform as
\begin{equation}
\begin{aligned}
\Re C & \rightarrow +\Re C ,\\
\Im C & \rightarrow -\Im C  .
\end{aligned}
\end{equation}
Using these parity properties we can now draw an analogy between the boundary graviton $C$ and the scalar potential $\phi$ in electrodynamics, whose real and imaginary components define the electric and magnetic fields, respectively, and transform in a similar way (see table \ref{EManalogy} and references \cite{Henneaux:2018cst,Henneaux:2018gfi,Henneaux:2018hdj} for related works on the parity odd component of the metric).
The standard \emph{electric boundary condition} \eqref{StandradBC} therefore sets to zero the magnetic component of the metric while the alternative \emph{magnetic boundary condition} \eqref{orthogonalBC} sets to zero its electric component.
More generally, we can impose the \emph{dyonic boundary condition} \eqref{NewBC} which implies that both the real and imaginary modes of $C$ are turned on.
Our main goal in this paper is to explore the physics of the imaginary part of $C$.

\begin{table}[]
	\vspace{0 mm}
	\centering
	\renewcommand{\arraystretch}{2.5}
	\begin{tabular}{ccc}
		\toprule[1.5pt]
		\textbf{Gravity}              & 																							 & \textbf{QED}						 		 \\ 	[2ex] \midrule[1pt]
$
\begin{aligned}
\\
\Re C  &\sim 	\Big[	D_{\zb}^2 C_{zz} +D_z^2 C_{\zb\zb} 	\Big]_{\mI^+_{-}}  \\ \\
\Im C &\sim 
\Big[
D_{\zb}^2 C_{zz} -D_z^2 C_{\zb\zb}
\Big]_{\mI^+_{-}}
\end{aligned}
$
	    									        &    	          																	       &  
$
\begin{aligned}
\\
\Re \phi  &\sim   \Big[      \pazb A_z^{(0)}+\paz A_{\zb}^{(0)}\Big]_{\mI^+_{-}} 
\\[10pt]
\Im \phi  &\sim     \Big[\pazb A_z^{(0)}-\paz A_{\zb}^{(0)}\Big]_{\mI^+_{-}}
\end{aligned}
$	                    				    	    \\ 	[8ex] \hline
$
\begin{aligned}
\\
T(f)  \\[30pt]
\mM(f)  
\end{aligned}
$
&    	          																	       &  
$
\begin{aligned}
\\
Q_{\text{electric}}(\epsilon) &= \frac{1}{e^2}\int_{\mI^+_-}d^2 z  \gammaflat \, \epsilon \,  F_{ru}^{(2)}
\\[10pt]
Q_{\text{magnetic}}(\epsilon) &=  \frac{i}{2\pi} \int_{\mI^+_-} d^2 z  \, \epsilon \,  F_{z\zb}^{(0)}
\end{aligned}
$	                    				    	    \\ 	[8ex]
		\bottomrule[1pt]
	\end{tabular}
	\vspace{1.5 cm}
	\caption{The analogy between gravity and electrodynamics.
		Here $\phi$ is the scalar potential in electrodynamics, $F_{ru}^{(2)}$ and $F_{z\zb}^{(0)}$ are the leading terms in the asymptotic expansions of the electric and magnetic fields and we have used the gauge condition $A_u^{(0)}=0$ (see \cite{Strominger:2015bla,Strominger:2017zoo} for more details).
		The real and imaginary parts of the boundary graviton describe the parity even and parity odd components of the metric, respectively, in analogy with the electric and magnetic fields. Correspondingly, the analogy extends to the charges that are associated with each field component.
	}
	\label{EManalogy}
\end{table}

In electrodynamics, each one of the field components (electric and magnetic) define a conserved charge.
Similarly in gravity, the real and imaginary parts of the boundary graviton are associated with two different conservation laws.
The conserved charges are defined in terms of the complex Weyl scalar $\Psi_2$, whose leading component in the asymptotic expansion is given by
\begin{equation}\label{PsiC}
\begin{aligned}
\Psi_2^0(u,z,\zb) & \equiv - \lim_{r\rightarrow \infty} \left(r C_{uzr\zb} \gammaflatt\right) \\
&= - m_B +\frac{1}{4} C^{zz}N_{zz} +\frac{1}{4} (\gammaflatt)^2 \Big( D_{\zb}^2 C_{zz}-D_z^2 C_{\zb\zb}\Big).
\end{aligned}
\end{equation}
The real part of $\Psi^0_2$ defines the BMS supertranslation charge
\begin{equation}\label{superT}
\begin{aligned}
T(f) &= - \frac{1}{4\pi G}
\int_{\mI^+_-}  d^2 z \gammaflat f(z,\zb) \Big[ \Re \Psi_2^0(u,z,\zb) \Big]_{\mI^+_-} \\
&= \frac{1}{4\pi G}
\int_{\mI^+_-} d^2 z \gammaflat f(z,\zb) m_B.
\end{aligned}
\end{equation}
In section \ref{supertsection} we review BMS supertranslations and show how they are related to the standard electric boundary condition.
The imaginary part of $\Psi^0_2$ defines an independent charge
\begin{equation}\label{NewCharge}
\begin{aligned}
\mM(f)   &= - \frac{1}{4\pi G}
\int  d^2 z \gammaflat f(z,\zb) \Big[   \Im \Psi_2^0(u,z,\zb) \Big] _{\mI^+_-}
\\
&=  \frac{i}{16\pi G}
\int_{\mI^+_-}  d^2 z \,  \gammaflatt f(z,\zb) 
\Big(
D_{\zb}^2 C_{zz} -D_z^2 C_{\zb\zb}
\Big).
\end{aligned}
\end{equation}
In the literature $\mM (f)$ is sometimes referred to as the \emph{gravitational magnetic aspect} or \emph{gravitomagnetic monopole}. We will refer to $\mM (f)$ as the \emph{dual supertranslation} charge, in reference to the electric-magnetic duality (see table \ref{EManalogy}). Clearly, the dual supertranslation charge vanishes when the standard electric boundary condition \eqref{StandradBC} is imposed, but will acquire a non-zero value for the magnetic or dyonic boundary conditions.
As we will see soon, the dual supertranslation charge is conserved identically and therefore it is completely fixed by the boundary conditions.

The expansion of the metric around past null infinity $\mI^-$ takes the form
\begin{equation}
\begin{aligned}
ds^2 &= -dv^2 +2dv dr+2r^2 \gammaflat dz d\zb \\
&+
\frac{2m_B^-}{r} dv^2 +r D_{zz}^- dz^2 +r D_{\zb\zb} d\zb ^2 -2V_z dv dz -2 V_{\zb} dv d\zb \\
&+ \frac{1}{r}\left(
\frac{4}{3}\left(N_z^- +v  \paz m_B^-  \right) - \frac{1}{4} \paz \left(D_{zz} D^{zz}\right)
\right) dv dz + c.c.
\\
&+ \dots ,
\end{aligned}
\end{equation}
with $v$ the advanced time, 
\begin{equation}
V_z = -  \frac{1}{2} D^z D_{zz}
\end{equation}
and
\begin{equation}
N_{zz}^{-} = \pa_u D_{zz} .
\end{equation}
Following the previous discussion at $\mI^+$, the new dyonic boundary conditions on the radiative modes at $\mI^-$ will take the form
\begin{equation}\label{NewBCD}
-i \alpha \Big[
D_{\zb}^2 D_{zz} -D_z^2 D_{\zb\zb}
\Big]_{\mI^-_{+}}
+\beta 
\Big[
D_{\zb}^2 D_{zz} +D_z^2 D_{\zb\zb}
\Big]_{\mI^-_{+}}
=0,
\end{equation}
\begin{equation}
N_{zz}\eval  _{\mI^-_{+}}=0,
\end{equation}
which are solved by
\begin{equation}\label{boundaryGravitonD}
D_{zz} \eval _{\mI^-_{+}} = D_z^2 D (z,\zb),
\end{equation}
where $D(z,\zb)$ is a \emph{complex} function.
Similarly, the supertranslation and dual supertranslation charges can be defined at $\mI^-$
\begin{equation}
\begin{aligned}
T^-(f) &= \frac{1}{4\pi G}
\int_{\mI^-_+}  d^2 z \gammaflat f(z,\zb) m_B^-  \\
\mM^- (f)  &= -  \frac{i}{16\pi G}
\int_{\mI^-_+}  d^2 z \,  \gammaflatt f(z,\zb) 
\Big(
D_{\zb}^2 D_{zz} -D_z^2 D_{\zb\zb}
\Big).
\end{aligned}
\end{equation}
Note that the dual supertranslation charge, as well as the parity odd part of \eqref{NewBCD}, contain a relative minus sign with respect to their $\mI^+$ counterparts.
The reason is that the sphere's coordinates on $\mI^-$ are antipodally mapped to the ones on $\mI^+$.
The antipodal map reverse the orientation of the sphere and therefore flips the sign of parity odd fields.

The scattering problem in general relativity is solved by a map of the Cauchy data on $\mI^-$ to that on $\mI^+$.
Generally speaking, this map is obtained by evolving the Cauchy data using the Einstein equations.
However, the equations of motion should be supplemented with boundary conditions to make 
the scattering problem well defined.
In \cite{Strominger:2013jfa}, the following matching conditions between the boundary data for the parity even components of the metric on $\mI^-_+$ and $\mI^+_-$ were proposed
\begin{equation}\label{matchEven1}
m_B (z,\zb) \eval  _{\mI^+_-}   = m_B^- (z,\zb)  \eval  _{\mI^-_+}.
\end{equation}
\begin{equation}\label{matchEven2}
\Re C (z,\zb) \eval _{\mI^+_-} = \Re D (z,\zb) \eval _{\mI^-_+},
\end{equation}
In the same spirit, here we propose the following matching condition for the parity odd metric component
\begin{equation}\label{matchOdd}
\Im C (z,\zb) \eval _{\mI^+_-} =  - \Im D (z,\zb) \eval_{\mI^-_+}.
\end{equation}
These matching conditions, together with the boundary conditions \eqref{NewBC} and \eqref{NewBCD}, are all invariant under Lorentz and CPT transformations.

\begin{figure}[]
	\begin{center}
	\includegraphics[scale=1]{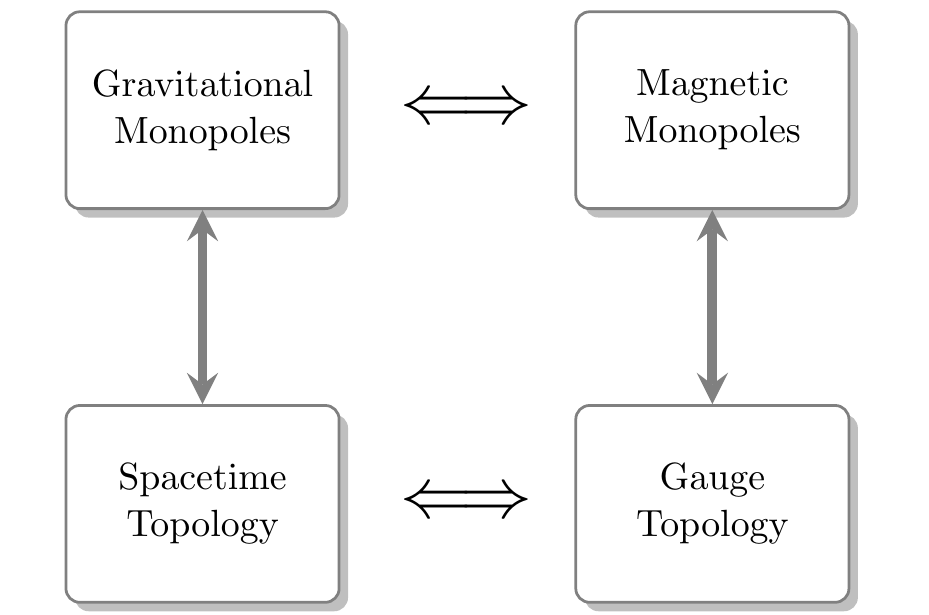}
\end{center}
	\caption{The analogy between our work and the theory of magnetic monopoles. For brevity, we refer to dual supertranslation charged objects simply as \emph{gravitational monopoles}, which are analogous to magnetic monopoles in electrodynamics. In 
	the same way that the magnetic monopole charge partitions the space of gauge fields into distinct topological sectors, the 
	gravitational monopole charge partitions the space of spacetime metrics into topologically distinct sectors.
	}
	\label{Analogy}
\end{figure}

We would like to emphasize a key difference between the matching conditions of the boundary graviton \eqref{matchEven2}-\eqref{matchOdd} and the one of the Bondi mass \eqref{matchEven1}.
Matching of the boundary graviton follows from Lorentz invariance \cite{Strominger:2013jfa}, but it is not enough to ensure stress-energy conservation, which we need to impose explicitly by the matching condition on the Bondi mass.
We refer to the former as the \emph{trivial} matching conditions, while the latter is imposed explicitly.
To emphasize the importance of this difference, let us resort again to the analogy with electrodynamics, where the matching condition of the scalar potential follows from Lorentz invariance \cite{He:2014cra}. In electrodynamics this is enough to ensure magnetic charge conservation (in the absence of magnetic sources), but electric charge conservation has to be imposed explicitly. In other words, magnetic charge is conserved \emph{identically} and follows from continuity of the potential, while electric charge conservation is imposed explicitly. Similarly in gravity, we see that the dual supertranslation charge is conserved identically due to continuity of the boundary graviton, while supertranslation conservation is a direct result of stress-energy conservation (which is analogous to charge conservation in QED). 
In QED, despite the absence of magnetic charges, a non-trivial solution that carries magnetic charge still exists and is known as the \emph{Dirac string} \cite{Dirac:1931kp}.
This magnetic configuration is possible due to its non-trivial topology in gauge space. The magnetic charge then classifies the gauge field into distinct topological sectors.
In gravity, the identical conservation of the dual supertranslation charge therefore suggests that it may describe a configuration with a non-trivial \emph{spacetime topology}. 
Indeed, we will show in section \ref{NewSection} that the global dual supertranslation charge $\mM_{\text{global}}$ is a 
topological invariant of spacetime by proving that it is independent of the metric.
This property distinguishes dual supertranslations from other asymptotic charges that appear in the literature, like BMS supertranslations and superrotations.
The analogy between the topological structure in gravity and in electrodynamics, due to the presence of monopoles, is described in figure \ref{Analogy}.

The simplest case of a configuration with a non-zero dual supertranslation charge is when $\Im \Psi_2^0$ is a constant
\begin{equation}\label{TBcharge}
\Im \Psi_2^0 =  \ell .
\end{equation}
The charge defined by $\Im \Psi_2^0$ is then known as the NUT charge, and $\ell$ is called the NUT parameter.
The Taub-NUT metric is an example of a solution that obeys equation \eqref{TBcharge}.
On the other hand, the Taub-NUT metric is known to exhibit an infinite string singularity similar to the Dirac string.
The presence of a cosmic string singularity implies that the celestial space is an infinite-fold covering of the two sphere with a branch cut at the location of the sphere.
This example shows the connection between dual supertranslations and the topology of the celestial space.
We will study this example in detail.

The paper is organized as follows. In section \ref{supertsection} we review BMS supertranslations and show that the new dyonic boundary condition we introduced in \eqref{NewBC} is consistent with their action on the phase space.
In section \ref{rotSection} we study new aspects of BMS superrotations that follow from the new boundary conditions we introduced. We discuss a puzzle that was not addressed in the literature so far - a mismatch between the action of the superrotation charge on the phase space and the corresponding action of the Lie derivative. This puzzle resembles the one that was encountered in \cite{He:2014laa} for supertranslations, and we solve it in a similar manner. The new boundary conditions are
 essential to this discussion; though they are more general than~(\ref{StandradBC},\ref{gravRadCond}), they are strong enough to 
 resolve the puzzle.
In section \ref{NewSection} we study the new topological symmetry that is associated with the dual supertranslation charges \eqref{NewCharge} and which we name \emph{dual supertranslations}. We show that the charge $\mM$ is composed of soft gravitons only, compute its action on the phase space and show that it cannot be reproduced by diffeomorphisms.
Moreover, we prove that the global dual supertranslation charge $\mM_{\text{global}}\equiv \mM(f)\eval_{f=1}$ is independent of any metric deformation and is therefore a topological invariant of spacetime.
We study the Taub-NUT metric as an example of a solution that carries a dual supertranslation charge and an imaginary boundary graviton, and use it to demonstrate the resulting topological structure.
In order to understand how the new degrees of freedom, that were introduced by the new boundary conditions, arise in a physical process, in section \ref{MemSec} we study the classical effect of a stress tensor on the phase space and the corresponding symmetry operation.
The effect associated with supertranslations goes under the name of \emph{the memory effect}.
We derive new formulas for the effect that is associated with superrotations and dual supertranslations.
We show that there is no memory in these cases and that the effect cannot be reproduced by regular diffeomorphisms like in the case of supertranslations.
In section \ref{ScatteringSec} we study a vacuum solution of the Einstein equations that interpolates between regions of spacetime with different dual supertranslation charge.
The solution describes the scattering of two impulsive gravitational plane waves.
The dual supertranslation charge vanishes before the instant of collision and acquires a non-zero value after the two plane waves collide.
This solution can be interpreted as the formation of a cosmic string with NUT charge and provides a physical interpretation for dual supertranslations.
We end with a few comments and topics for future study of the new topological symmetry.

 \section{Review of BMS Supertranslations}\label{supertsection}
 
 We start by reviewing the action of supertranslations on the phase space and on spacetime, the puzzle that was encountered in \cite{He:2014laa} concerning the Dirac brackets and the proposed solution using the boundary condition \eqref{StandradBC}. We show that BMS supertranslations are independent of the new degree of freedom introduced by the dyonic boundary condition \eqref{NewBC}, which therefore does not alter the analysis of \cite{He:2014laa}.

The generator of BMS supertranslations on $\mI^+$, which is given by \eqref{superT}, can be decomposed into two parts
\begin{equation}
T(f) = \Tsoft (f) + \Thard (f)
\end{equation}
using the $uu$ component of the Einstein equations $G_{uu} = 8\pi T^M_{uu}$
\begin{equation}\label{uuEquation}
\begin{aligned}
\pa_u m_B &= \frac{1}{4} \left(D_z^2 N^{zz}+D_{\zb}^2 N^{\zb\zb}\right) - T_{uu} ,\\
T_{uu} &=  4\pi G \lim_{r \rightarrow \infty } \left(r^2 T^M_{uu}\right) + \frac{1}{4} N_{zz}N^{zz}.
\end{aligned}
\end{equation}
The hard part of the generator is given by
\begin{equation}
\Thard (f) = \frac{1}{4 \pi G} \int _{\mI^+} du d^2z f(z,\zb) \gammaflat T_{uu}
\end{equation}
while the soft part is
\begin{equation}
\begin{aligned}\label{tsoft}
\Tsoft(f) &= \frac{1}{8\pi G} \int_{\mI^+} du\,  d^2z  \, \pa_u \Big(  \pazb U_{z}  +   \paz U_{\zb}  \Big) f(z,\zb) \\
&=
- \frac{1}{16\pi G} \int_{\mI^+}  du \,  d^2z \, \gammaflatt  \,  \Big(  D^2_{\zb} N_{zz}+ D^2_{z} N_{\zb \zb}    \Big)  f(z,\zb) .
\end{aligned}
\end{equation}

The action of supertranslations on the metric is described by the vector field
\begin{equation}
\xi_f = f \pa_u-\frac{1}{r} \left(D^z f \pa_z +D^{\zb} f \pa_{\zb} \right)
+D^z D_z f \pa_r
+\mO \left( r^{-2} \right),
\qquad
f= f(z,\zb).
\end{equation}
In particular, it implies that the radiative data transforms as
\begin{equation}
\mL_f C_{zz} =f \pa_u C_{zz}-2D_z^2 f .
\end{equation}
In \cite{He:2014laa} it was noticed that using the canonical Dirac bracket 
 \begin{equation}\label{Cbracket}
 \{N_{\zb\zb}(u,z,\zb),C_{ww}(u',w,\wb)\} = 16\pi G \delta(u-u') \delta^2(z-w) \gammaflat,
 \end{equation}
the action of the supertranslation charge
 \begin{equation}\label{wrongCom}
 \{T(f),C_{zz}\} = f \pa_u C_{zz}-D_z^2 f \neq \mL_f C_{zz}
 \end{equation}
 is inconsistent with the variation of the metric under supertranslations. The inhomogeneous term in \eqref{wrongCom} is off by a factor of 2. Reference~\cite{He:2014laa} proposed to resolve this inconsistency by imposing the boundary condition \eqref{StandradBC}. The boundary condition \eqref{StandradBC} relates the two polarization modes of the zero-momentum gravitons and as a result both terms in \eqref{tsoft} contribute to the bracket $ \{T(f),C_{zz}\}$, in such a manner 
 that the contribution to the inhomogeneous part is doubled. In the rest of this section we will review this construction and show that the more general boundary condition we introduced in \eqref{NewBC} is also consistent with the action of BMS supertranslations on the phase space.
 
 In addition to the boundary field $C(z,\zb)$, ref.~\cite{He:2014laa} introduced the boundary field $N(z,\zb)$, defined by
 \begin{equation}\label{defN}
 \int_{-\infty}^{\infty} du N_{zz} (u,z,\zb)= D_z^2 N(z,\zb) .
 \end{equation}
 The standard boundary condition \eqref{StandradBC} implies that both $C$ and $N$ are real functions while the new dyonic boundary condition \eqref{NewBC} implies that they are complex. However, the soft part of the supertranslations generator is a function of the real part of $N$ only
 \begin{equation}\label{TsoftReN}
 \Tsoft (f) = - \frac{1}{8 \pi G} \int d^2z \, \gammaflatt \, f(z,\zb) \, D_z^2 D_{\zb}^2     \Big(  \Re N(z,\zb )\Big)  .
 \end{equation}
BMS supertranslations are therefore completely independent of the magnetic part of the metric and the analysis of \cite{He:2014laa} follows through also when the more general dyonic boundary condition \eqref{NewBC} is imposed.

 By continuity, and using \eqref{Cbracket}, ref.~\cite{He:2014laa} found the algebra of the boundary fields
 \begin{equation}\label{ps1}
 \begin{aligned}
  \{C_{\zb\zb}(u,z,\zb),C_{ww}(u',w,\wb)\} &= 8\pi G \Theta(u-u') \delta^2(z-w) \gammaflat, \\
 \{ \Re C(z,\zb),C_{ww}(u',w,\wb)\}  & = -8 GD_w^2 \left(S \ln |z-w|^2\right),  \\
\{ \Re N(z,\zb),C_{ww}(u',w,\wb)\}  & =  16 GD_w^2 \left(S \ln |z-w|^2\right),  \\
\{ \Re N(z,\zb), \Re C(w,\wb)\}   &=  16 G  S \ln |z-w|^2 ,
 \end{aligned}
 \end{equation} 
  where $\Theta(x)=\text{sign}(x)$ while the function
 \begin{equation}\label{Sfunc}
 S = \frac{(z-w)(\zb-\wb)}{(1+z\zb)(1+w\wb)}
 \end{equation}
 obeys
 \begin{equation}
 \begin{aligned}
 D_w^2  \left(S \ln |z-w|^2\right) &= \frac{S}{(z-w)^2}, \\
 D_{\zb}^2 D_w^2  \left(S \ln |z-w|^2\right) &=  \pi \gammaflat \delta^2(z-w).
 \end{aligned}
 \end{equation}
 The resulting action of the BMS generator on the set of fields that span the phase space is then given by
 \begin{equation}\label{ps2}
 \begin{aligned}
 \{T(f) , N_{zz }\} &= f \pa_u N_{zz}, \\
  \{T(f) , C_{zz }\} &= f \pa_u C_{zz} -2D_z^2 f, \\
   \{T(f) , \Re N\} &= 0, \\
      \{T(f) , \Re C\} & = -2f.
 \end{aligned}
 \end{equation}
 In particular, $  \{T(f) , C_{zz }\} = \mL_f C_{zz}$ and the puzzle is resolved.
This review summarizes the discussion about the electric sector of the phase space which is described by the BMS supertranslations generator and the real part of the boundary fields.
In the following sections we will study the magnetic counterpart and the interplay between the two sectors.

 \section{New Properties of BMS Superrotations}\label{rotSection}
 
 We now turn to study BMS superrotations in light of the new dyonic boundary conditions that we introduced in \eqref{NewBC}. We will see that these boundary conditions are essential in solving a puzzle similar to the one encountered for supertranslations in \eqref{wrongCom}.
 
 The generator of BMS superrotations on $\mI^+$ \cite{Barnich:2011mi,Barnich:2016lyg}
 \begin{equation}
 Q(Y) = \frac{1}{8\pi G} \int _{\mI^+_-} d^2z \left(Y_{\zb}N_z + Y_z N_{\zb}\right)
 \end{equation}
 can be decomposed into soft and hard parts
 \begin{equation}
 Q(Y) = Q_S(Y) +Q_H(Y) 
 \end{equation}
 using the $uz$ component of the Einstein equations $G_{uz} = 8\pi T^M_{uz}$
 \begin{equation}\label{uzEquation}
 \begin{aligned}
 \pa_u N_z &= \frac{1}{4}\paz  \left(D_z^2 C^{zz} - D_{\zb}^2 C^{\zb\zb}\right)  -u \pa_u \paz m_B  - T_{uz} ,\\
 T_{uz} &= 8 \pi G \lim_{r \rightarrow \infty } \left(r^2 T^M_{uz}\right) - \frac{1}{4} \paz \left( C_{zz}N^{zz}\right) -\frac{1}{2}C_{zz}D_z N^{zz}.
 \end{aligned}
 \end{equation}
 The hard part is given by
   \begin{equation}
 Q_{H}(Y) =  \frac{1}{8\pi G}
 \int _{\mI^+} du d^2z \left(
 Y_{\zb} T_{uz} + Y_z T_{u \zb} +u \pa_z Y_{\zb} T_{uu}+u \pa_{\zb} Y_{z} T_{uu}
 \right) ,
 \end{equation}
 while the soft part is
  \begin{equation}\label{QYsoft}
 Q_{S}(Y) = 	-\frac{1}{16\pi G} \int _{\mI^+} du d^2z \gammaflatt u
\left( D_z ^3 Y^z N_{ \zb \zb } +  D_{\zb} ^3 Y^{\zb} N_{ z z }\right).
 \end{equation}

The action of superrotations on the metric is described by the vector field
\begin{equation}\label{xiY}
\xi_Y = 
(1+\frac{u}{2r})Y^z \paz 
-\frac{u}{2r} D^{\zb}D_zY^z \pazb
-\frac{1}{2}(u+r)D_z Y^z \pa_r
+\frac{u}{2} D_z Y^z \pa_u+c.c. .
\end{equation}
At leading order, the metric component $g_{\zb\zb}$ transforms under \eqref{xiY} as
\begin{equation}
\mL_Y g_{\zb\zb} = 2r^2 \gammaflat \pazb Y^z + \mO(r) .
\end{equation}
In order to preserve the falloff conditions, the vector field $Y^z$ on the two sphere is therefore required to obey
\begin{equation}\label{rotationCond}
\pa_{\zb} Y^z =0.
\end{equation}
 Locally, this is solved by
 \begin{equation}
 Y^z = \{  z^n , \, i \, z^n \} ,
 \end{equation}
 for any integer $n$. However, only the choices $n=0,1,2$ lead to a globally defined vector fields, which are holomorphic functions on the sphere. These are the six global conformal Killing vector fields on $S^2$. For a general integer $n$ the vector fields are meromorphic functions on the sphere and the symmetry algebra is promoted to the infinite-dimensional algebra of
local conformal transformations. In this case the falloff conditions on the metric are violated at isolated points. For example, if $Y^z = \frac{1}{z-w}$, then $\pazb Y^z = 2\pi \delta^2(z-w) \neq 0$ so that the falloff condition is violated at $z=w$. The rest of the metric components transform in a way that preserves the falloff conditions.
 
 At the next order, the same metric component transforms as
  \begin{equation}
 \delta_Y C_{zz} = \frac{u}{2} D \cdot Y N_{zz} 
 +Y \cdot D C_{zz}
 -\frac{1}{2} D \cdot Y C_{zz} 
 +2 D_z Y^z C_{zz}
 -u D_z^2 \left(D \cdot Y \right) ,
 \end{equation}
where
\begin{equation}
D \cdot Y \equiv D_z Y^z + D_{\zb} Y^{\zb} .
\end{equation}
 On the other hand, using the Dirac bracket \eqref{Cbracket} we can compute the action of the superrotation charge on $C_{zz}$
 \begin{equation}
 \begin{aligned}
 \{Q_S(Y),C_{zz}\} &= -u D_z^3 Y^z ,  \\
 \{Q_H(Y),C_{zz}\} &= \frac{u}{2} D \cdot Y N_{zz} 
 +Y \cdot D C_{zz}
 -\frac{1}{2} D \cdot Y C_{zz} 
 +2 D_z Y^z C_{zz}.
 \end{aligned}
 \end{equation}
 We now encounter a mismatch similar to the puzzle with supertranslations
 \begin{equation}\label{puzzle2}
 \{Q(Y),C_{zz}\} \neq \delta_Y C_{zz}.
 \end{equation}
 The mismatch, again, is in the inhomogeneous term. This problem does not seem to have been addressed in the literature.
 
 Here we suggest a solution in the spirit of \cite{He:2014laa}. Let us define the boundary field $\Nt (z,\zb)$
 \begin{equation}
\int_{-\infty}^{\infty} du \, u \,  N_{zz} (u,z,\zb) = D_z^2 \Nt  (z,\zb) .
\end{equation}
 Here, $\Nt$ is a complex function. In fact, the imaginary part of $\Nt$ contributes to the soft superrotation charge, as can be seen by rewriting \eqref{QYsoft} as
 \begin{equation}
 \begin{aligned}
 Q_S(Y) &= Q_S^{\Re} (Y) + Q_S^{\Im} (Y),\\
 Q_S^{\Re} (Y)  &= + \frac{1}{16\pi G} \int_{\mI^+_-} d^2 z\, \gamma^{z\zb} 
\Big[  D_z^2 D_{\zb}^2 \left( D \cdot Y \right)\Big] \Re \Nt (z,\zb),
 \\
 Q_S^{\Im} (Y)  &= 
 -\frac{i}{16\pi G} \int_{\mI^+_-}   d^2 z \,  \gamma^{z\zb} 
 \Big[D_z^2 D_{\zb}^2 \left( D_z Y^z - D_{\zb}Y^{\zb} \right)\Big] \Im \Nt (z,\zb).
 \end{aligned}
 \end{equation}
We have identified the contributions to $Q_S(Y)$ due to the real and imaginary parts of $\Nt$ as $Q_S^{\Re} (Y)$ and $Q_S^{\Im} (Y)$, respectively.

 The action of $\Nt$ on $C_{zz}$ can be deduced from
 \begin{equation}
 \begin{aligned}
 D_z^2 \{    \tilde{N} (z,\zb), C_{\wb \wb} (u,w, \wb) \} &= 
 \int _{- \infty }^{\infty} du' \, u'  \{   N_{zz} (u', z,\zb), C_{\wb \wb} (u,w,\wb) \}
 \\
 &=
 -16 \pi G u \gammaflat \delta^2(z-w).
 \end{aligned}
 \end{equation}
 The above equation is solved by
 \begin{equation}\label{NtBracket}
\{  \Nt (z,\zb), C_{\wb \wb} (u,w,\wb) \} = -16 G u D_{\wb}^2 \left( S \log |z-w|^2\right),
 \end{equation}
from which we can also deduce
 \begin{equation}
\{  \Nt (z,\zb), C(w, \wb) \} = -16 G u  S \log |z-w|^2 .
\end{equation}
 Let us emphasize again that both $\Nt$ and $C$ are in general complex functions.
 Each one of them, therefore, contain two real scalar degrees of freedom and we have to determine the action of each one of them on the phase space separately.

\emph{A priori} we don't know how the Dirac bracket \eqref{NtBracket} is decomposed in terms of the real and imaginary parts of $\Nt$. The idea is to determine this decomposition by the requirement that 
 \begin{equation}\label{correctBracket}
\{Q(Y), C_{zz}\} =   \delta_Y C_{zz}.
\end{equation}
To demonstrate this logic we use the following ansatz 
 \begin{equation}\label{NtBracketDecompose}
 \begin{aligned}
\{    \ReNt , C_{\wb \wb} (u,w,\wb)  \}  &= - a_1 \,  16 G  u D_{\wb}^2 \left( S \log |z-w|^2\right), \\
\{  i\,  \ImNt, C_{\wb \wb} (u,w,\wb) \}  &= - a_2 \,  16 G u D_{\wb}^2 \left( S \log |z-w|^2\right),
 \end{aligned}
 \end{equation}
 where
 \begin{equation}
a_1 + a_2 =1
 \end{equation}
 such that the sum of the two equations in \eqref{NtBracketDecompose} reproduces the Dirac bracket \eqref{NtBracket}. We can now compute the action of $Q_S(Y)$ on $C_{zz}$
 \begin{equation}
 \{ Q_S(Y),C_{zz}\} = - u D_z^2 \Big[  a_1 D \cdot Y - a_2 \left(D_z Y^z -D_{\zb} Y^{\zb}\right) \Big].
 \end{equation}
The requirement that equation \eqref{correctBracket} is obeyed now determines the coefficients $a_1$ and $a_2$ to be
\begin{equation}
a_1 =1 , \qquad a_2 =0.
\end{equation}
 In other words, we have to impose that $\ImNt$ commutes with $C_{zz}$
  \begin{equation}\label{ImNBracket}
 \{    \ImNtz, C_{\wb \wb} (u,w,\wb) \}  =0 ,
 \end{equation}
and that the action of $\ReNt$ on the radiative data is given by
  \begin{equation}\label{ReNBracket}
 \{    \ReNtz  , C_{\wb \wb} (u,w,\wb)  \}  = -  16 G  u D_{\wb}^2 \left( S \log |z-w|^2\right).
 \end{equation}
We conclude that if and only if we impose the Dirac brackets \eqref{ImNBracket}-\eqref{ReNBracket}, the puzzle \eqref{puzzle2} is resolved. We would like to emphasize that this conclusion does not depend on the ansatz \eqref{NtBracketDecompose}, which we used for demonstration purpose only.

 Imposing \eqref{ImNBracket} and \eqref{ReNBracket} we can now compute the action of $Q_S^{\Im}(Y)$ on the phase space
 \begin{equation}
 \{Q_S^{\Im}(Y),C_{zz}\}= \{Q_S^{\Im}(Y),C\} = \{Q_S^{\Im}(Y),N_{zz}\}=\{Q_S^{\Im}(Y),N\} =0.
 \end{equation}
 Namely, $Q_S^{\Im}(Y)$ leaves the entire phase space spanned by $\{C_{zz},N_{zz},C,N\}$ invariant! This is a result of the requirement that equation \eqref{correctBracket} be satisfied.
 In particular, it implies that $Q_S^{\Im}(Y)$ commutes with both soft and hard supertranslations
 \begin{equation}
 \{Q_S^{\Im}(Y) , \Tsoft (f)\}= \{Q_S^{\Im}(Y) , T_H(f)\}=0.
 \end{equation}
 Let us also note that 
 \begin{equation}
 \{\ReNtz , \Im \Nt (w,\wb) \} =0,
 \end{equation}
 which implies
 \begin{equation}
 \{   Q_S^{\Re} (Y) , Q_S^{\Im} (Y') \} =0.
 \end{equation}
Finally, let us mention that $Q_S^{\Im}$ does not commute with the hard superrotations charge $Q_H$.

 \section{Properties of Dual BMS Supertranslations}\label{NewSection}
 
 We would like to explore now the consequences of the magnetic boundary condition \eqref{orthogonalBC}, which implies that $C(z,\zb)$ and $N(z,\zb)$ are imaginary functions on the complex plane.
 The soft supertranslation charge vanishes in this case, but the dual supertranslation charge doesn't.

We start by showing that $\mM(f)$ is a soft charge.
 Using the expression for $C_{zz}$ in terms of the gravitons' Fourier modes \cite{Choi:2017bna}
 \begin{equation}
 C_{zz}(u,z,\zb)= -\frac{i \kappa}{8 \pi ^2} \gammaflat
 \int _0 ^{\infty} d \omegak \left[a_+(\omegak\hat{\bold{x}}_z)  e^{-i \omegak u} -a_-^{\dagger}(\omegak\hat{\bold{x}}_z)  e^{i \omegak u}   \right] ,
 \end{equation}
the dual supertranslation charge \eqref{NewCharge} can be brought to the form
 \begin{equation}
 \mM (f) = 
 \lim_{\omegak \rightarrow 0}
 \frac{i \omegak}{4\pi \kappa} \int d^2 z 
\Big[
 \left(  a_+(\omegak\hat{\bold{x}}_z)   +a_-^{\dagger}(\omegak\hat{\bold{x}}_z)  \right) D_{\zb}^2 f
 - \text{h.c.}
\Big],
 \end{equation}
 which shows that $\mM(f)$ receives contributions only from soft graviton modes.
 As opposed to the soft supertranslation charge, $\mM(f)$ is conserved by itself and does not have a hard counterpart.

We now follow the same route that was taken in section \ref{supertsection} for supertranslations and apply it to the new charge in order to compute its action on the phase space.
First, we rewrite \eqref{NewCharge} in terms of the boundary field $N$\footnote{Here we are dropping a term proportional to $\Big[
	D_{\zb}^2 C_{zz} -D_z^2 C_{\zb\zb}
	\Big]_{\mI^+_+}$ for simplicity. This term could be important for certain applications, but is not needed for our purposes.}
 \begin{equation}\label{TsoftImN}
\begin{aligned}
\mM(f) &=  \frac{-i}{16\pi G}
\int_{\mI^+} du  d^2 z \,  \gammaflatt f(z,\zb) 
\Big(
D_{\zb}^2 N_{zz} -D_z^2 N_{\zb\zb}
\Big)
\\
&=
\frac{-i}{16\pi G}
\int_{\mI^+} du  d^2 z \,  \gammaflatt
\Big[
N_{zz}   D_{\zb}^2 f(z,\zb)   -  N_{\zb\zb}  D_z^2f(z,\zb)   
\Big]
\\
&= \frac{1}{8 \pi G} \int d^2z \, \gammaflatt \, f(z,\zb) \, D_z^2 D_{\zb}^2     \Big[\Im N(z,\zb )\Big]
\\
&= \frac{1}{8 \pi G} \int d^2z \, \gammaflatt \, D_z^2 D_{\zb}^2     \left[  f(z,\zb)  \right]     \Im N(z,\zb )
 .
\end{aligned}
\end{equation}
 The algebra that follows from \eqref{Cbracket} is given by
  \begin{equation}
 \begin{aligned}
 \{C_{\zb\zb}(u,z,\zb),C_{ww}(u',w,\wb)\} &= 8\pi G \Theta(u-u') \delta^2(z-w) \gammaflat, \\
 \{   \Im C(z,\zb),C_{ww}(u',w,\wb)\}  & = +8 G i \, D_w^2 \left(S \ln |z-w|^2\right),  \\
 \{   \Im N(z,\zb),C_{ww}(u',w,\wb)\}  & =  - 16 G i \, D_w^2 \left(S \ln |z-w|^2\right),  \\
 \{   \Im N(z,\zb),   \Im C(w,\wb)\}   &=  - 16 G  S \ln |z-w|^2 .
 \end{aligned}
 \end{equation} 
 Together with \eqref{TsoftImN}, it implies that the action of $\mM (f)$ on the phase space is
  \begin{equation}\label{actionNew}
 \begin{aligned}
 \{\mM (f) , N_{zz }\} &= 0, \\
 \{\mM (f) , C_{zz }\} &=  -2 i D_z^2 f, \\
 \{\mM (f) , \Im N\} &= 0, \\
 \{\mM (f) , \Im C\} & = -2 f.
 \end{aligned}
 \end{equation}
This algebra looks very much like the ``imaginary" counterpart of soft supertranslations.
The question that follows immediately is whether there exists a spacetime diffeomorphism that corresponds to the action of $\mM (f)$ on the phase space \eqref{actionNew}. As we will now show, the answer to this question is \emph{no}.

A spacetime diffeomorphism is described by a vector field $\xi^{\mu}(u,r,z,\zb)$. We can expand $\xi^{\mu}$ around $\mI^+$ as follows
\begin{equation}\label{diffExp}
\xi^{\mu} = \sum_{n=0}^{\infty} \frac{1}{r^n} \xi^{\mu(n)}(u,z,\zb).
\end{equation}
We will focus our attention on the transformation of the following metric components under the above diffeomorphism
\begin{equation}
\begin{aligned}
\mL_{\xi} g_{uu} & =  \dots+
\frac{2\xi^{u(0)}  \pa_{u}  m_B+\dots}{r}
+\dots , \\
\mL_{\xi} g_{rz} & =  -\gammaflat \Big(   \xi^{\zb (1)}  +D^{\zb} \xi^{u(0)}  \Big) +\dots ,\\
\mL_{\xi} C_{zz} & = 
\xi^{u(0)}  \pa_{u} C_{zz} + 2 D_z \xi^{(1)}_z.
\end{aligned}
\end{equation}
To reproduce the action of $\mM (f)$ on the phase space we would like to set the component $\xi^{u(0)}$, which generates hard transformations, to zero. On the other hand, the falloff conditions on the metric, which ensure asymptotic flatness, require that $g_{rz}=0$. The falloff conditions, together with the requirement that $\xi^{u(0)}$ vanishes, then sets $\xi^{z(1)}$ and $\xi^{\zb(1)}$ to zero, therefore rendering the action of the transformation on the phase space trivial. In other words, there is no way to generate a non-trivial homogeneous transformation on the phase space while setting the non-homogeneous terms to zero and obeying the falloff conditions at the same time, using diffeomorphisms.
Spacetime diffeomorphisms therefore \emph{cannot} reproduce the action of $\mM (f)$ on the phase space.

Let us make a couple of remarks on possible generalizations of our analysis. First, one could relax the falloff conditions on the metric to allow for isolated singularities on the sphere, in a similar way to superrotation transformations. However, it is easy to see that even if we allow for isolated singularities to appear in the $g_{rz}$ component of the metric, it is impossible to generate a non-trivial homogeneous term in $C_{zz}$. The second remark is that one could possibly include linear terms in the expansion \eqref{diffExp}, like the ones that appear in superrotation transformations. However, that will not change our analysis, which depend only on the components $\xi^{u(0)}$, $\xi^{z(1)}$ and $\xi^{\zb(1)}$.

A simpler way to understand our result is to notice that we would need a diffeomorphism that acts nontrivially
on the asymptotic metric while leaving invariant all matter fields (since there is no hard component in $\mM (f)$). This is
clearly impossible.

We have shown that the dual supertranslation charge $\mM (f)$, for any function $f(z,\zb)$ on the sphere cannot be 
a spacetime diffeomorphisms. This is consistent with the fact that our boundary conditions set to zero supertranslations, which
are the only diffeomorphism acting nontrivially on the asymptotic fields. In
the next subsection we will show that even when a nonzero $T (f)$ is allowed by the dyonic boundary conditions, it still
commutes with $\mM (f)$ so that $\mM (f)$ is still invariant under all diffeomorphisms.
The global dual supertranslation charge, given by
\begin{equation}\label{TotalS}
\mM_{\text{global}} = \mM(f) \eval _{f=1} = \frac{i}{16 \pi G} \int_{\mI^+_-} d^2 z \Big(   \pazb \pa^z C_{zz} - \paz \pa^{\zb} C_{\zb \zb}  \Big),
\end{equation}
obeys a stronger condition - it is invariant under \emph{any} deformation of the metric that is globally well defined on the two sphere.
Under a general deformation of the metric
\begin{equation}
C_{zz} \rightarrow C_{zz} + a_{zz},
\end{equation}
where $a_{zz}$ is a 2-form, the charge $\mM_{\text{global}}$ does not change because the integrand in \eqref{TotalS} is a total derivative. In other words, $\mM_{\text{global}}$ itself vanishes unless there are some defects on the sphere that prevents the metric from being a globally-defined 2-form, in which case it is given by
\begin{equation}\
\mM_{\text{global}}  = \frac{i}{16 \pi G} \oint_{\mC} \Big(  dz \,  \pa^z C_{zz} - d \zb \, \pa^{\zb} C_{\zb\zb}   \Big),
\end{equation}
where $\mC$ is a contour around the defect.
We conclude that $\mM_{\text{global}} $ is invariant under any smooth deformation of the metric and therefore classifies the geometry of the asymptotic space into distinct topological sectors.

 \subsection*{Dyonic Boundary Conditions}

 We now consider the general boundary condition \eqref{NewBC}, which implies that both $C$ and $N$ are complex function, and compute the commutation relations between the real and imaginary sectors. The fundamental bracket
 \begin{equation}
 \{C_{\zb\zb}(u,z,\zb),C_{ww}(u',w,\wb)\} = 8\pi G \Theta(u-u') \delta^2(z-w) \gammaflat
 \end{equation}
 implies that
 \begin{equation}\label{NstarC}
 \{   N^* (z,\zb),C_{ww}(u',w,\wb)\}   =  - 16 G  \, D_w^2 \left(S \ln |z-w|^2\right).
 \end{equation}
 Further acting with the operator $\int du' \pa_{u'}$ on \eqref{NstarC} we find
 \begin{equation}
 \{   N^* (z,\zb), N(w,\wb)   \}   =  0 . 
 \end{equation}
 Namely, the real and imaginary parts of $N$ commute
  \begin{equation}
 \{   \Re N (z,\zb),  \Im N(w,\wb)   \}   =  0 ,
 \end{equation}
 which in turn implies that dual supertranslations commute with supertranslations
 \begin{equation}
 \{ T(f_1), \mM (f_2 ) \} =0.
 \end{equation}
 This result agrees with the fact that the Bondi mass does not change under dual supertranslations. Finally, let us note that dual supertranslations do not commute with superrotations\footnote{In fact, dual supertranslations commute with the soft part of the superrotation charge, but not with its hard part.}.

 \subsection*{Example}

 An example of a solution that admits a non-zero dual supertranslation charge is the Taub-NUT metric, which is given by
 \begin{equation}\label{TaubNUTMetric}
 ds^2 = 
 -f(r) \left(dt +2 \ell \cos \theta d \varphi\right)^2
 +\frac{1}{f(r)} dr^2
 +\left(r^2 + \ell^2 \right)\left(d\theta^2 +\sin ^2 \theta d\varphi ^2\right)
 \end{equation}
 with
 \begin{equation}
 f(r) = \frac{r^2 -2 m r -\ell^2}{r^2+\ell^2}.
 \end{equation}
Here $m$ is the mass of the black hole and $\ell$ is called the NUT parameter. 
The Taub-NUT metric \eqref{TaubNUTMetric} has two horizons located at
\begin{equation}
r_{\pm} = m \pm \sigma ,
\end{equation}
where
\begin{equation}
 \sigma  = \sqrt{ m^2 +\ell ^2 } .
\end{equation}
The domain $r_- < r < r_+$ describes the Taub region while the domains $r>r_+$ and $r< r_-$ describe the NUT region.

The complex Weyl scalar $\Psi_2$ of the Taub-NUT solution is given by
\begin{equation}\label{Psi2TN}
\Psi_2 = - \frac{m- i \ell}{(r+i \ell)^3}
\end{equation}
 and its leading component in the asymptotic expansion is
 \begin{equation}
 \Psi_2^0 = -m + i \ell .
 \end{equation}
 As we mentioned in the introduction, the imaginary part of $ \Psi_2^0$ is given by the NUT parameter and therefore the dual supertranslation charge is non-zero
 \begin{equation}\label{TNcharge}
 \mM (f) =  \frac{\ell}{4 \pi G} \int d^2 z \gammaflat \, f(z,\zb).
 \end{equation}
The global dual supertranslation charge is therefore 
 \begin{equation}
\mM _{\text{global}}=  \frac{\ell}{ G}.
\end{equation}

The Taub-NUT metric \eqref{TaubNUTMetric} is locally isomorphic to flat space everywhere except at $\theta=0$ and $\theta=\pi$ where there is a conical singularity. This line singularity is the gravitational analogue of the Dirac string for magnetic monopoles. In order to study the asymptotic structure of the Taub-NUT spacetime we will bring it to the Bondi form \eqref{futureMetric}.
To do this we first write it in the retarded system of coordinates and we use complex variables on the two sphere
\begin{equation}
ds^2 = 
-f(r)
\left(
du+dr+ i \ell  \frac{1-|z|^2}{|z|^2} \frac{z d\zb -\zb dz}{1+ |z|^2} 
\right)^2
+\frac{1}{f(r)}dr^2
+2(r^2 + \ell^2) \gammaflat dz d\zb
\end{equation}
where
\begin{equation}
z= \tan \left( \frac{\theta}{2} \right) e^{i \varphi} .
\end{equation}
In order to bring the metric into the Bondi form, we have to remove the $g_{rr}$ component using
\begin{equation}\label{uChange}
\begin{aligned}
u  \rightarrow u +V(r) , \qquad \qquad
V'(r) = - \frac{f(r) \pm 1  }{f(r)},
\end{aligned}
\end{equation} 
and then employ the following change of variables on the two sphere
\begin{equation}\label{changeZ1}
\begin{aligned}
z &\rightarrow z -i   \frac{1-|z|^4}{2   \zb }  \frac{\ell}{r} ,\\
\zb &\rightarrow \zb + i   \frac{1-|z|^4}{2   z }  \frac{\ell}{r}.
\end{aligned}
\end{equation}
Note that \eqref{changeZ1} is not defined at either $z=0$ or $z=\infty$ (corresponding to $\theta=0$ and $\theta=\pi$ respectively).
The result is that the Taub-NUT metric can be brought to the Bondi form \eqref{futureMetric} with
\begin{equation}
C_{zz} = -  i \ell \gammaflat \frac{1+|z|^4}{z^2}
\end{equation}
everywhere except at the location of the string. This implies that the boundary graviton is given by
\begin{equation}\label{ImCTN}
C (z,\zb)=   -  2 i \ell      \log  \frac{(1+|z|^2)^2}{|z|^2}  .
\end{equation}
We conclude that the Taub-NUT solution serves as an example in which the boundary graviton is purely imaginary, namely - it follows from the new magnetic boundary condition that we introduced in \eqref{orthogonalBC}.
As it was discussed above, the imaginary part of the boundary graviton can develop a non-zero value only when the topology of the asymptotic spacetime is non-trivial.
Indeed, the presence of an isolated string singularity in the Taub-NUT solution implies that the celestial space is 
 a covering of the sphere branched at the location of the string.

The infinite string singularity can be partly eliminated by a change of variables. This procedure results in the appearance of closed timelike curves and is regarded as pathological in the literature. In this paper we do not attempt to address the question of how to eliminate the string in order to achieve a singularity-free monopole solution. Instead, we will treat the infinite string as a physical object. However, for the sake of completeness we will briefly review the partial de-singularization
procedure and the resulting asymptotic structure.
It is possible to remove the singularity of \eqref{TaubNUTMetric} at $\theta=0$ by the change of coordinates $t =t'-2 \ell \varphi$, which brings the metric to the form
 \begin{equation}\label{SemiMetric}
 \begin{aligned}
 ds^2 &= 
-f(r) \left(dt' -4 \ell \sin ^2 \frac{\theta}{2} d \varphi\right)^2
+\frac{1}{f(r)} dr^2
+\left(r^2 + \ell^2 \right)\left(d\theta^2 +\sin ^2 \theta d\varphi ^2\right)\\
&=
 -f(r)
\left(
du'+dr-2 i \ell \frac{z d\zb -\zb dz}{1+ |z|^2  } 
\right)^2
+\frac{1}{f(r)}dr^2
+2(r^2 + \ell^2) \gammaflat dz d\zb.
 \end{aligned}
 \end{equation}
Using \eqref{uChange} and the following change of variables on the sphere
  \begin{equation}\label{changeZ}
 \begin{aligned}
 z &\rightarrow z +i \ell \frac{z(1+|z|^2)}{r},\\
 \zb &\rightarrow \zb -i \ell \frac{\zb(1+|z|^2)}{r} ,
 \end{aligned}
 \end{equation}
 the metric is brought to the Bondi form everywhere except at the location of the semi-infinite string ($z=\infty$ or equivalently $\theta=\pi$) with
 \begin{equation}
 C_{zz} = - 2i \ell \gammaflat \zb ^2.
 \end{equation}
This implies that the boundary graviton is given by
 \begin{equation}\label{ImCTN}
 C (z,\zb)= - 4 i \ell \log (1+|z|^2).
 \end{equation}

 \section{The Memory Effect}\label{MemSec}

The new boundary conditions \eqref{NewBC} have introduced two new boundary fields - the imaginary parts of $C(z,\zb)$ and $N(z,\zb)$ - to the phase space. As it was shown in the previous section, the imaginary part of $N$ is related to the dual supertranslation charge, which, in turn, acts non-trivially on the imaginary part of $C$. But how exactly $\Im C$ arises in a physical process?
To answer this question, in this section we will consider the classical effect of a stress tensor on the asymptotic phase space of the theory.

Before discussing the new degrees of freedom, we will first briefly review the classical effect of a stress tensor on the real part of the boundary graviton.
Consider, for example, a burst of radiation described by a stress tensor $T_{uu}$ that is non-zero only in some finite interval $u_i<u<u_f$.
The radiation flux will induce a non-zero change in $\Re C$ which can be computed using the $uu$ component of the Einstein equations \eqref{uuEquation} 
\begin{equation}\label{dReC}
\begin{aligned}
\Delta \Re C (z,\zb) &\equiv  \Re C (z,\zb)   \eval _{\mI^+_+}   -   \Re C (z,\zb)   \eval _{\mI^+_-} \\
&= \frac{2}{\pi}
\int d^2 w \gamma_{w\wb} G(z,w) \left( \int_{-\infty}^{+\infty} du \, T_{uu} (u,w,\wb)   +\Delta m_B \right) ,
\end{aligned}
\end{equation}
 where $G(z,w)$ is the Green's function
\begin{equation}
\begin{aligned}
G(z,w) &= S \log S ,\\
D_z^2 D_{\zb}^2 G &= \pi \gammaflat \delta^2 (z-w) ,
\end{aligned}
\end{equation}
defined in terms of the function $S(z,w)$ defined in \eqref{Sfunc}. This process goes under the name of \emph{the memory 
effect} and it shows how a radiation burst can interpolate between two distinct vacua that are related by a BMS supertranslation 
\cite{Strominger:2014pwa}. We would like to emphasize that only the real part of $C$ appears in the equation of motion 
\eqref{uuEquation} and therefore the memory effect described by the formula \eqref{dReC} applies only to it and not to the 
imaginary part of $C$. Another example for the classical effect of a stress tensor on the asymptotic phase space was studied in 
\cite{Hawking:2016sgy}, where it was shown how to implant supertranslation hair on a Schwarzchild black hole using a shock 
wave. The formula \eqref{dReC} applies in this case too, as well as in any classical process, since it is simply an inversion of the
 equation of motion.

In a similar way, BMS superrotations transform $\Re C$ by a term linear in $u$.
In other words, they map two distinct states that are characterized by different values of $\pa_u \Re C$.
A natural question to ask is whether there exists a memory effect in the boundary graviton to leading order 
in the asymptotic expansion associated with superrotations. Subleading memory effects for superrotations were studied 
in~\cite{comp} and a discussion about the subject can be found in~\cite{Pasterski:2015tva}, but the subject does not 
seem to have been fully addressed in the literature.
The answer to this question can be inferred, again, from equation \eqref{uuEquation}, which dictates the relation between $\pa_u \Re C$ and the stress tensor 
 \begin{equation}\label{duReC}
 \pa_u  \Re C \eval _{\mI^+_{\pm}} = \frac{2}{\pi}
 \int d^2 w \gamma_{w\wb} G(z,w) \,  \Big[  T_{uu} (u,w,\wb)     \Big]_{\mI^+_{\pm}}  .
 \end{equation}
 Note that the only way $\pa_u  \Re C$ can acquire a non-zero value is by having a non-vanishing stress tensor at $u=\pm \infty$.
This situation is very different from the memory effect for $\Delta \Re C $, where due to the $u$-integral the effect is \emph{integrated} and a burst of radiation at some finite interval will induce a finite memory. For superrotations we see that the equivalent process will yield an instantaneous effect rather than an \emph{integrated} memory effect.
 Note also that this effect is due to the real part of superrotation only (as defined in section \ref{rotSection}). The imaginary part of superrotations leaves the phase space invariant and does not induce any change in the physical parameters.
 
 Following the same logic we will now consider the effect of a stress tensor on $\Im C$. To do this we use the $uz$ component of the Einstein equation \eqref{uzEquation}. Inverting this equation we derive the following formula
 \begin{equation}\label{dImC}
 \begin{aligned}
 \Im C (z,\zb) \eval _{\mI^+_{\pm}}  &= - \frac{i}{\pi ^2}
 \int d^2 q \,  \gamma^{q \qb} \,  G(z,q)
 \int d^2 w \,
 \pa_{\wb} \log |q-w|^2
\,   \times
 \\
 & 
\qquad \qquad \qquad \qquad 
\Big[ 
\pa_u N_w(u,w,\wb) - u \pa_u \pa_w m_B(u,w,\wb) -T_{uw} (u,w,\wb)
\Big]_{\mI^+_{\pm}},
 \end{aligned}
 \end{equation}
 where we have used \eqref{boundaryGraviton} and $\pa_z \pa_{\wb} \log |z-w|^2 = -2\pi \delta^{(2)}(z-w)$.
 The expression \eqref{dImC} is more convoluted than the previous formulas, but here again we see that it does not involve an integral over the null direction $u$, similarly to the formula \eqref{duReC} for $ \pa_u  \Re C $. We therefore conclude that there is \emph{no} integrated memory effect for $\Im C$. Nonetheless, there is an instantaneous effect similar to the effect of superrotations.
 Let us emphasize, however, that while superrotations induce a non-zero Bondi news, dual supertranslations do not.
 The violation of asymptotic flatness in this case is therefore milder in the sense that it does not require an infinite source of energy.
 
Following the same route, one could think of a similar effect for $\pa_u \Im C \sim \pa_u T_{uz}$. This kind of effect will require a linearly diverging stress tensor and does not correspond to any of the symmetry transformations we discussed in this paper.
 
  \begin{table}[]
 	\vspace{0 mm}
 	\centering
 	\renewcommand{\arraystretch}{2.5}
 	\newcolumntype{C}[1]{>{\centering\arraybackslash}p{#1}}
 	\begin{tabular}{ccC{2.8cm}c}
 		\toprule[1.5pt]
 		\textbf{Symmetry}              & \textbf{Phase Space}																							 & \textbf{Memory}						 &				 \textbf{Diffeomorphism}				 \\ 
 		[2ex] \midrule[1pt]
 		Supertranslations               &    $\Delta \Re C  \sim         \int _{-\infty}^{\infty} du \, T_{uu}$		                    &   (Integrated) memory	     &                Regular diffeomorphism 		         \\ 
 		[2ex] 
 		Superrotations                    &    $ \pa_u \Re C \eval  _{\mI^+_{\pm}}   \sim     T_{uu}	 \eval _{\mI^+_{\pm}}$       		     &   No memory          				    &                Singular diffeomorphism     		    \\ 
 		[2ex] 
 		Dual Supertranslations      &   $ \Im C	\eval _{\mI^+_{\pm}}  \sim    T_{uz}			\eval  _{\mI^+_{\pm}} 	$             	    &    No memory           				   &                 No diffeomorphism      				  \\ 
 		[2ex]\bottomrule[1pt]
 	\end{tabular}
 	\vspace{0.4 cm}
 	\caption{The classical effect of a stress tensor on the phase space and its relation to the three symmetry operations.}
 	\label{MemoryTable}
 \end{table}

Finally, let us emphasize that these results are consistent with the spacetime transformations that are associated with the different symmetries.
The effect of supertranslations can be reproduced by a regular spacetime diffeomorphism, or equivalently by a stress tensor.
On the other hand, there is no \emph{regular} diffeomorphism that can reproduce the action of superrotations nor dual supertranslations, therefore rendering a corresponding memory effect impossible.
After all, the stress tensor measures the response of the metric to small changes.
We summarize the classical effect of the stress tensor on the phase space, its relation to diffeomorphisms and to the three symmetry transformations in table \ref{MemoryTable}.

 \section{Scattering of Two Impulsive Gravitational Plane Waves}\label{ScatteringSec}
 
 In the previous section we described how the stress tensor can implant a non-zero imaginary component of the boundary graviton. However, this component can also arise and play an important role also in vacuum solutions. In this section we 
 will describe such a solution of the Einstein's vacuum equations, that involves a scattering of two impulsive gravitational 
 plane waves. This scattering process interpolates between regions of spacetime with different dual supertranslation 
 charge and different values of the field $\Im C$. It therefore provides an explicit realization of dual supertranslations.

\begin{figure}[]
	\begin{center}
	\includegraphics[scale=1.3]{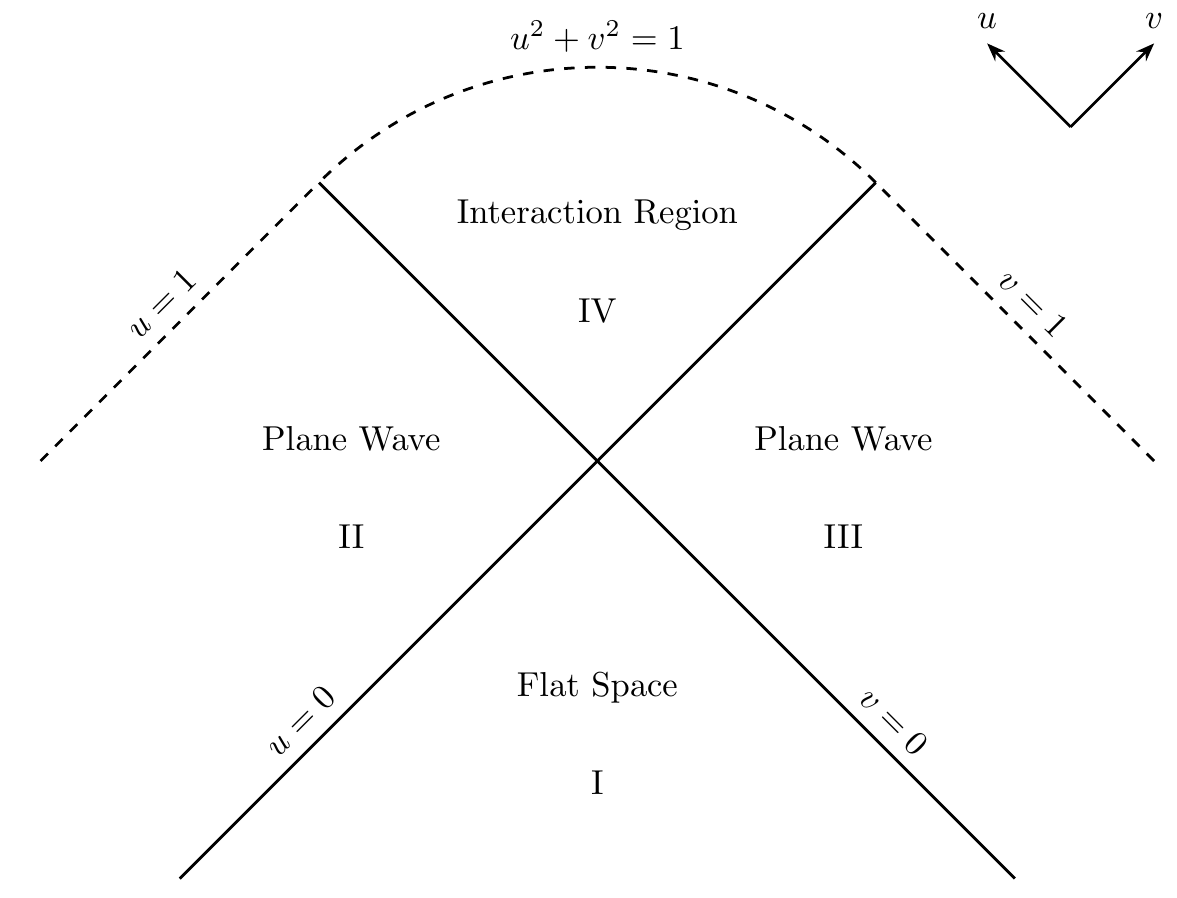}
\end{center}
	\vspace{1 cm}
 	\caption{The Penrose diagram of the scattering process of two impulsive gravitational plane waves. The two waves propagate along $u=0$ and $v=0$, respectively and travel freely until the instant of collision. The interaction region is locally isomorphic to the Taub region of the Taub-NUT metric and can be extended beyond the even horizon which is described by the dashed line.
 	}
 	\label{fig:scattering}
 \end{figure}

 The solution was found by Ferrari and Iba\~nez \cite{Ferrari:1988nu} (see also \cite{Ferrari:1987cs,Ferrari:1987yk}), and is based on a method due to Penrose and Khan \cite{Khan:1971vh} for constructing solutions that describe scattering of two impulsive gravitational plane waves.
The solution is given by the following metric
 \begin{equation}\label{FIsol}
 ds^2 =
- 16A \frac{X}{F} du dv +s^2 \frac{Y}{X}\left(  dx - 2q \mu dy  \right)^2 +\frac{X}{Y} dy^2 .
 \end{equation}
 Here $u$ and $v$ are two null coordinates\footnote{Our notation of the null coordinates is related to that of \cite{Ferrari:1988nu} by $u\longleftrightarrow v$.}, $A$ is a constant representing the amplitude of the waves and we defined
 \begin{equation}
\begin{aligned}
F &= 2 \frac{\eta^2 -\mu ^2 }{\sqrt{1-s-j}\sqrt{1-s+j}} , \\
X&= 1+ \eta^2 +2p \eta, \qquad
Y= \frac{1}{1-\mu ^2 }, \\
\eta &= \frac{1}{2} \sqrt{(1+j)^2-s^2}+\frac{1}{2} \sqrt{(1-j)^2-s^2},\\
\mu &= \frac{1}{2} \sqrt{(1+j)^2-s^2}-\frac{1}{2} \sqrt{(1-j)^2-s^2} ,\\
 s&= 1- \ut ^2 - \vt^2 , \qquad
  j=  \vt^2 -\ut ^2 ,
\end{aligned}
 \end{equation}
 where
 \begin{equation}
 \ut = u H(u), \qquad \vt = v H(v).
 \end{equation}
 $H(u)$ is the Heaviside step function that describes the impulsive burst of the plane waves. The parameter $p$ represent the angle between the direction of the polarization of the two colliding waves. It takes the values $-1<p<1$. When $p=\pm 1$ the two waves have parallel polarizations. The parameter $q$ is related to $p$ by $p^2 +q^2=1$. The Penrose diagram of the metric \eqref{FIsol} is depicted in figure \ref{fig:scattering}.

 The metric \eqref{FIsol} interpolates between three different regions of spacetime.
 The first and simplest is the region $u,v<0$, where the metric \eqref{FIsol} reduces to that of flat spacetime
 \begin{equation}
 ds^2 = - 4A du dv +dx^2 +dy^2.
 \end{equation}
 This is the region of spacetime before the collision of the two plane waves at $u=0,v=0$.
The second region is $u\geq 0,v<0$, where the metric takes the following form
\begin{equation}\label{PW1}
\begin{aligned}
ds^2 &= -\frac{ 4A (1+u^2 +2pu)}{\sqrt{1-u^2}} du dv
\\
&+
\frac{1-u^2}{1+u^2 +2pu} (dx+2qu dy)^2
+(1-u^2)(1+u^2 +2pu) dy^2.
\end{aligned}
\end{equation}
This metric describes a plane wave propagating freely in the $u$-direction.
At $u=1$ there is a coordinate singularity, whose nature we will discuss soon.
In the region $v\geq 0,u<0$ the metric is given by the same expression \eqref{PW1} with $u$ replaced by $v$ and $q$ by $-q$; it describes a plane wave propagating freely in the $v$-direction.

Finally, the most interesting region is the interaction region $u,v \geq 0$. It describes spacetime after the collision of the two plane waves. In this region the solution is locally isomorphic to the Taub region of the Taub-NUT metric. This can be seen using a change of variables between the coordinates $\left(u,v,x,y\right)$ presented in this section and the system of coordinates $\left(t,r,\theta,\varphi \right)$ that appear in equation \eqref{TaubNUTMetric}. The null coordinates $u$ and $v$ are related to $r$ and $\theta$ by
 \begin{equation}\label{change1}
 \begin{aligned}
\eta &= v \sqrt{1-u^2} +u \sqrt{1-v^2} = \frac{r-m}{\sigma} ,\\
\mu &= v \sqrt{1-u^2} -u \sqrt{1-v^2} =  \cos \theta  \\
 \end{aligned}
 \end{equation}
 and the $x,y$ coordinates are related to $t$ and $\varphi$ by
  \begin{equation}
 \begin{aligned}
 x&=  t ,\\
  y &= \sigma \varphi ,\\
 \end{aligned}
 \end{equation}
 and the parameters of the solution \eqref{FIsol} are related to the Taub-NUT parameters by
  \begin{equation}
 \begin{aligned}
 p&= \frac{m}{\sigma}, \qquad 
 &q&=  \frac{\ell}{\sigma},
 \qquad A= \sigma^2 = m^2 +\ell ^2 .
 \end{aligned}
 \end{equation}
Under this change of coordinates and mapping of the parameters the solution \eqref{FIsol} in the interaction region takes the standard Taub-NUT form \eqref{TaubNUTMetric}.
In the solution of Ferrari and Iba\~nez, the interaction region is described by the domain $0<\eta<1$ and $-\eta<\mu<\eta$, which corresponds to the region $u>0,v>0$ and $u^2+v^2 \leq 1$.
$\eta$ measures the time from the collision, where $\eta=0$ (corresponding to $u=v=0$) is the instant of collision and at $\eta=1$ (corresponding to $u^2+v^2=1$) there is an event horizon.
It is therefore easy to see that in the interaction region $f(r)<0$ and the solution describes the Taub region of the Taub-NUT metric, where $t$ is a spacelike coordinate and $r$ is a timelike coordinate.

 The complex Weyl curvature scalars of the solution \eqref{FIsol} are given by
 \begin{equation}
 \begin{aligned}
 \Psi_{0} &=  G(u,v) H(v)  + G_0(u,v)  \delta (v) ,\\
  \Psi_{4} &=  G(u,v) H(u)  +   G_4(u,v) \delta (u),  \\
    \Psi_{2} &=   \frac{(p-i q)^2}{2A (1+p\eta-iq \eta)^3} H(u)  H(v), \\
 G(u,v) &= -\frac{3(p-i q)^2}{2A (1+p\eta-iq \eta)^3} \frac{\chi+1-2iq \mu}{\chi+1+2iq \mu}, \\
 \chi &= (1+p\eta +\eta^2) \sqrt{\frac{1-\mu^2}{1-\eta^2}},
 \end{aligned}
 \end{equation}
 where $\eta$ and $\mu$ are given in terms of $u$ and $v$ in \eqref{change1}.
$\Psi_0$ and $\Psi_4$ represent the ingoing and outgoing transverse gravitational wave parts of the field. $\Psi_2$ represents the Coulomb part of the field and it is different from zero only in the interaction region. In terms of the standard Taub-NUT coordinates it takes the form \eqref{Psi2TN}.
For more details about the Weyl curvature scalars and the Newman-Penrose formalism see references \cite{Newman:1961qr}-\cite{Adamo:2014baa}.
For brevity, we avoid writing down explicitly the coefficients of the delta function terms (they can be inferred from \cite{Ferrari:1988nu,Ferrari:1987cs,Ferrari:1987yk}).

\subsection*{The Interaction Region}

The interaction region of the solution was fully extended in \cite{Ferrari:1988nu} where it was shown to be isomorphic to the Taub-NUT metric. In particular, it was shown that the solution is smooth and that there are no curvature singularities besides the delta functions arising from the shockwaves and the string singularity.

To understand the motion of the plane waves in the interaction region we first look at the Taub-NUT metric near the horizon and for a fixed value of $\phi$
\begin{equation}\label{NHTN}
ds^2 = \frac{1}{\sqrt{2m r_+}} \left(-\tau dt^2 +\frac{d\tau^2}{\tau}\right) + d\theta ^2,
\qquad
\tau = \frac{r-r_+}{\sqrt{2mr_+}}.
\end{equation}
In the Taub region $\tau < 0$ and using the following change of variables
\begin{equation}
\begin{aligned}
U &= a \sqrt{-\tau} e^{\frac{t}{2}}, \\
V&= a \sqrt{-\tau} e^{-\frac{t}{2}},
\end{aligned}
\qquad
a^2 =\frac{4}{\sqrt{2mr_+}},
\end{equation}
the metric \eqref{NHTN} describes the Milne region of flat spacetime
\begin{equation}
ds^2 = - dU dV +d \theta ^2.
\end{equation}
The plane waves are described by the wave equation
\begin{equation}\label{waves}
\eta = \pm \mu .
\end{equation}
We will describe the propagation of the wave that is coming from region II and which corresponds to the upper sign in \eqref{waves}. In the Milne coordinates the wave equation reads
\begin{equation}
U V = - \frac{\sigma}{m r_+} \theta ^2.
\end{equation}
We further define
\begin{equation}
\begin{aligned}
X & =2 \sqrt{\frac{m r_+}{\sigma}} \left( U+V \right), \\
T & =2 \sqrt{\frac{m r_+}{\sigma}} \left( U-V \right), \\
\end{aligned}
\end{equation}
in terms of which the wave equation becomes
\begin{equation}
T^2 - X^2 -\theta ^2 = 0.
\end{equation}
We therefore conclude that the plane wave coming from region II propagates in the light cone of the spacetime point $U=V=\theta =0$. In a similar way one can show that the plane wave coming from region III propagate in the light cone of the spacetime point $U=V=0, \theta =\pi$. 

\begin{figure}[]
	\begin{center}
		\includegraphics[scale=0.9]{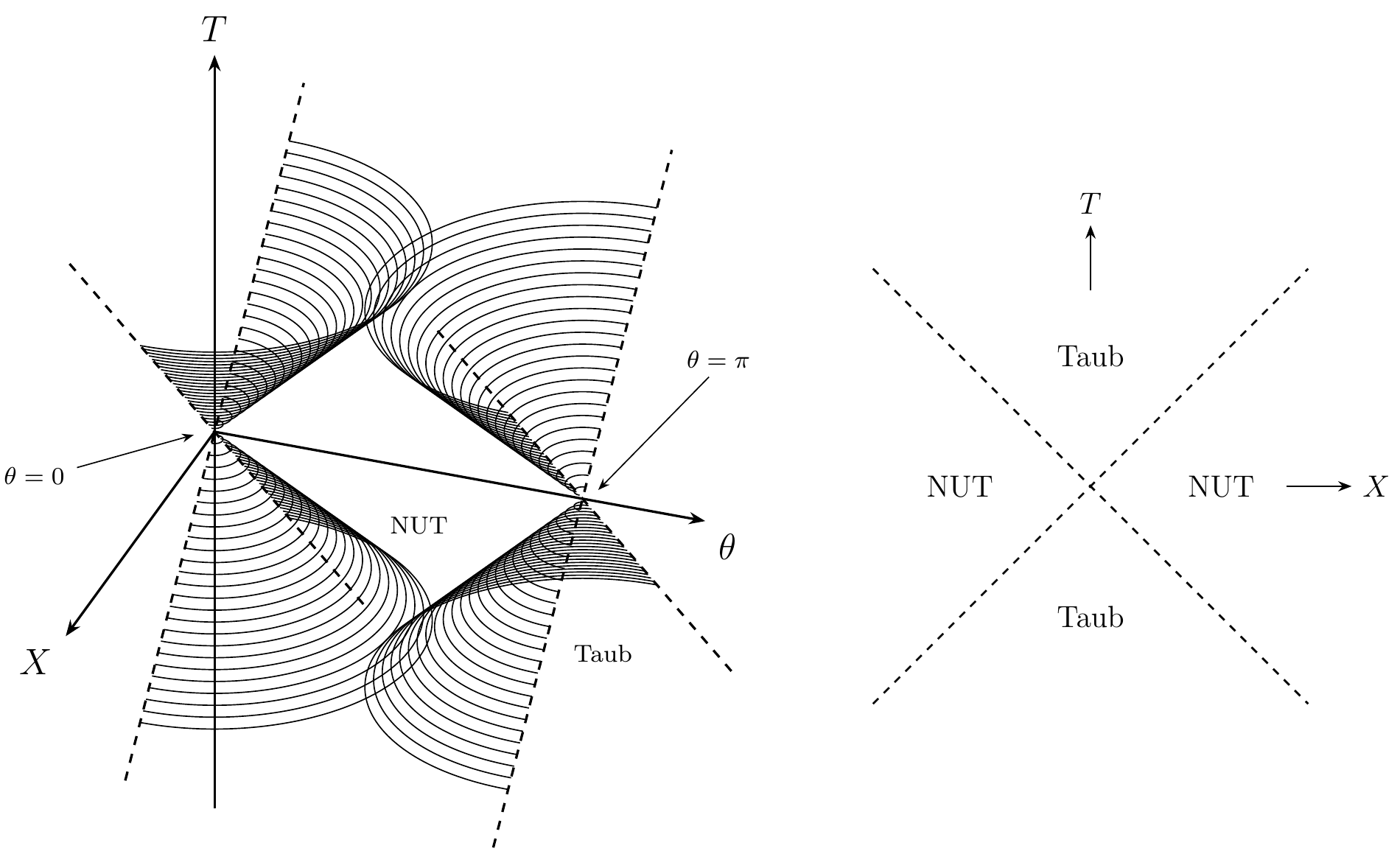}
	\end{center}
	\vspace{0.5 cm}
	\caption{Left: the interaction region projected onto the $T,X,\theta$ subspace. The two waves propagate inside the light cones of the two spacetime points $U=V=0, \, \theta =0 ,\pi$. Right: the interaction region projected onto the $T,X$ subspace.
	}
	\label{fig:Cones}
\end{figure}

The two waves propagate inside the light cones from the past Taub region into the future Taub region, as described in figure \ref{fig:Cones}, without ever passing through the NUT region.
The NUT region is static and can be written in terms of a cylindrical system of coordinates, also known as Weyl coordinates.
The advantage of this system is that it makes the physical picture of the string clear.
In these coordinates the NUT region takes the form
\begin{equation}\label{TNc}
ds^2 = - e^{2U} \left(dt +B d\phi \right)^2 +e^{-2U} \left[e^{2\gamma} \left(d\rho^2 +d x_3 ^2 \right) +\rho^2 d\phi ^2\right] ,
\end{equation}
where
\begin{equation}
\begin{aligned}\label{metricFun}
e^{2U} &= \frac{\left(R_+ +R_-\right)^2 - 4\sigma ^2  }{\left(R_++R_-+2m\right)^2 +4\ell^2}, \qquad
B=  \frac{\ell}{\sigma} \left( R_+ - R_-\right),
\\
e^{2\gamma} &= \frac{\left(R_+ +R_-\right)^2 - 4\sigma ^2 }{4 R_+ R_-} , 
\end{aligned}
\end{equation}
and
\begin{equation}
R^2_{\pm} = \rho^2 +\left( x_3 \pm \sigma  \right)^2.
\end{equation}
The relation to the spherical system of coordinates is given by
\begin{equation}\label{relations}
\rho = \sqrt{r^2 -2mr -\ell^2} \sin \theta , \qquad x_3 =(r-m) \cos \theta,
\end{equation}
which also implies
\begin{equation}\label{WeylFactors}
\begin{aligned}
e^{2U} &= f(r), \\
R_{\pm} &= r-m \pm \sigma \cos \theta, \\
B&= 2 \ell \cos \theta .
\end{aligned}
\end{equation}
Clearly this system of coordinates describes only the NUT region since the argument in the square root in \eqref{relations} has to be positive, and accordingly $\rho$ is a cylindrical radial coordinate that is defined in the domain $\rho>0$. In this region both $e^{2U}$ and $e^{2\gamma}$ are positive and therefore $t$ is a timelike coordinate and $\rho,x_3,\phi$ are spacelike. 
The conical singularity at $\theta = 0$ can be avoided by the change of coordinate $t \rightarrow t - 2 \ell \phi$.
The conical singularity at $\theta = \pi$ can be avoided by the change of coordinate $t \rightarrow t + 2 \ell \phi$.
We cannot avoid both conical singularities unless we allow for closed timelike curves. 
In this system of coordinates the two string segments lie on the $x_3$-axis in the regions $ |x_3| \geq \sigma$ and they are connected by the horizon which is located on the same axis at $ |x_3| < \sigma$
\begin{equation}
\begin{aligned}
&\text{half axis} &  	 	 \quad     &\theta =\pi 			&	  	\qquad  & 			\longleftrightarrow \qquad \rho =0,     \qquad 					x_3 \leq -\sigma , \\
&\text{horizon} &   	  \quad          &     r =  r_+ 		&		\qquad  & 				\longleftrightarrow \qquad \rho =0,         \;\; 				- \sigma < x_3 <+\sigma,  \\
&\text{half axis} &   		\quad	    &\theta = 0	           &			 \qquad  & 				\longleftrightarrow \qquad \rho =0,    \qquad			 +\sigma \leq x_3 .
\end{aligned}	
\end{equation}

\begin{figure}[]
	\begin{center}
		\includegraphics[scale=1.1]{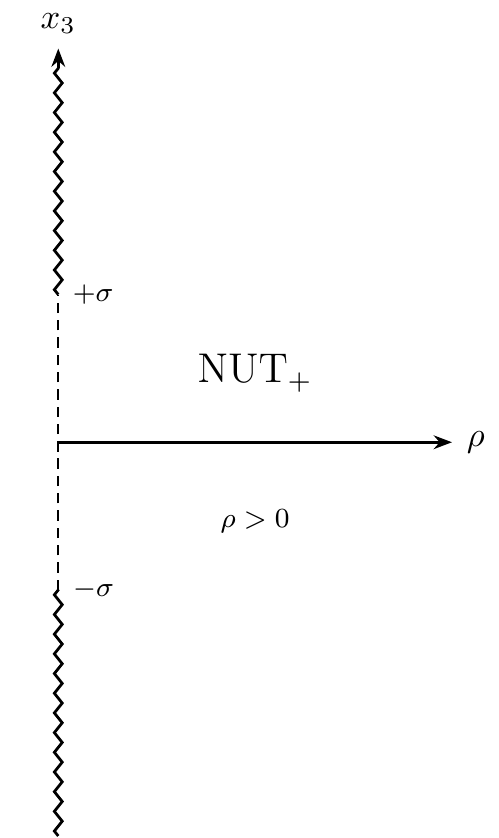}
	\end{center}
\vspace{-0.5cm}
	\caption{The NUT region of the Taub-NUT metric in the $x_3$-$\rho$ plane.
		The two string segments are stretched along the $x_3$-axis in the regions $|x_3| \geq \sigma$. The horizon at $r=r_+$, or equivalently $\rho=0$, lies in the domain $|x_3| \leq \sigma$ and is depicted by the dashed line.
	}
	\label{TNCylindrical}
\end{figure}

We can extend the metric \eqref{TNc} across the horizon into the Taub region using the following coordinates
\begin{equation}
\rho = - \sqrt{-(r^2 -2mr -\ell^2)} \sin \theta , \qquad x_3=(r-m) \cos \theta,
\end{equation}
under which the Taub-NUT metric \eqref{TaubNUTMetric} takes the form
\begin{equation}\label{TNc2}
ds^2 = - e^{2U} \left(dt +B d\phi \right)^2 +e^{-2U} \left[e^{2\gamma} \left( - d\rho^2 +dx_3^2 \right) -\rho^2 d\phi ^2\right] ,
\end{equation}
with the same functions $U,\gamma,B$ as in \eqref{metricFun} and
\begin{equation}
R^2_{\pm} = -\rho^2 +\left(x_3  \pm \sigma  \right)^2.
\end{equation}
The identities in equation \eqref{WeylFactors} remain the same in the Taub region.
Here both functions $e^{2U}$ and $e^{2\gamma}$ are negative and therefore $\rho$ is a timelike coordinate and $x_3,t,\phi$ are spacelike.
In this system of coordinates both the horizon and the string are at $\rho=0$ and $-\sigma <x_3 <+\sigma$. Note, however, that the horizon is a two dimensional surface in the full space and the string is one dimensional. This system of coordinates is not particularly useful to describe the Taub region and for future purposes we use it only to demonstrate the continuation across the horizon that separates the NUT and the Taub regions in Weyl coordinates. Formally the continuation is given by
\begin{equation}
\rho \rightarrow - i \rho.
\end{equation}

\subsection*{The Plane Wave Regions}

The plane wave regions II and III contain event horizons at $u=1$ and $v=1$, respectively.
The curvature remains finite on these horizons but geodesics never cross them and therefore they are not merely coordinates singularities.
They are known in the literature as "fold" singularities \cite{Matzner:1984pe} (see also section 8.3 of \cite{Griffiths:1991zp}).

To understand the fold singularities we will use Weyl coordinates to describe the metric in the plane wave regions.
Using the following system of coordinates
\begin{equation}
\begin{aligned}
x_3 &= -\sigma \left( u^2 +v^2 \right),  \qquad & \phi &= \frac{1}{\sigma } y\\
\rho &=  -\sigma \left( 1 -u^2 +v^2 \right),   \qquad & t&= x  ,
\end{aligned}
\end{equation}
the metric in region II \eqref{PW1} is brought to a form similar to \eqref{TNc2}
\begin{equation}\label{PWc}
ds^2 = - e^{2U} \left(dt +B d\phi \right)^2 +e^{-2U} \left[e^{2\gamma} \left(- d\rho^2 +dx_3^2 \right) - \left( \frac{x_3-\rho+\sigma}{2} \right)^2 d\phi ^2\right],
\end{equation}
with
\begin{equation}
\begin{aligned}\label{metricFun}
e^{2U} &= -  \frac{x_3-\rho+\sigma }{3\sigma -x_3+\rho  +2p \sqrt{2\sigma}\sqrt{\sigma -x_3+\rho  }   }  , \qquad
B=  2\ell \sqrt{-\frac{x_3-\rho-\sigma}{2\sigma}}  ,
\\
e^{2\gamma} &= - \sqrt{\frac{\sigma(x_3-\rho+\sigma)}{8(\sigma-x_3+\rho)(-\sigma-x_3-\rho)}} .
\end{aligned}
\end{equation}
In the entire region II both warp functions $e^{2U}$ and $e^{2\gamma}$ are negative. Therefore $\rho$ is a timelike coordinate and $x_3,t,\phi$ are spacelike coordinates. The metric \eqref{PWc} takes a form similar to the Taub metric in Weyl coordinates \eqref{TNc2}.
In the whole region $x_3<0, \rho<0$ and therefore the solution describes the lower half plane (or the southern hemisphere) and time is defined to be negative.
There is a coordinate singularity at $x_3-\rho+\sigma=0$ (corresponding to $u=1$) but the curvature remains finite there. However, the area of this horizon is zero and therefore it is a point in space. At this point $B=2\ell$ and there is a conical singularity similar to the one that appears in the Taub-NUT metric. This singularity can be interpreted as the endpoint of a semi-infinite string moving at the speed of light along its axis in the positive $x_3$ direction. At the point $x_3 = -\sigma,\rho=0$ (corresponding to $u=1,v=0$) the string's endpoint reaches the Taub-NUT horizon of the interaction region. 
Due to the presence of a fold singularity the metric cannot be extended beyond the horizon in the plane wave region.

\begin{figure}[]
	\begin{center}
		\includegraphics[scale=1.5]{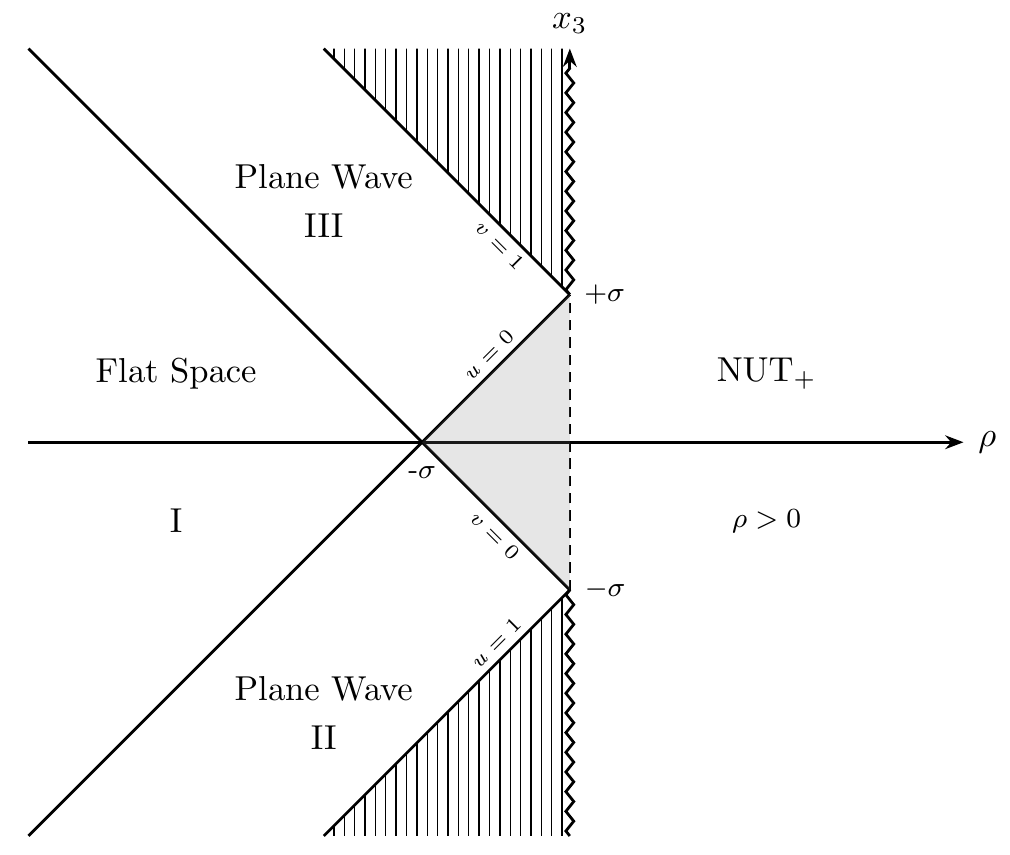}
	\end{center}
	\caption{The plane waves solution in the $x_3$-$\rho$ plane.
		The shockwaves at $u=0$ and $v=0$ correspond to $x_3-\rho-\sigma=0$ and $-x_3-\rho-\sigma=0$, respectively.
		The null lines at $u=1$ and $v=1$ correspond to $x_3-\rho+\sigma =0$ and $-x_3-\rho+\sigma =0$, respectively. On these lines there is a conical singularity and therefore the metric cannot be extended beyond them.
		The gray area is the interaction region IV (before the extension across the horizon), which is isomorphic to the part $m<r<r_+$ of the Taub space, and it can be extended beyond the horizon (depicted by the dashed line).
		Across the horizon the metric describes the NUT space, which includes two string segments (depicted by the zigzag lines) stretching from the horizon along the $x_3$-axis.
		In the NUT region $(\rho>0)$ the coordinate $\rho$ is spacelike, while in the region $\rho<0$ it is timelike.
		The two null lines $x_3-\rho+\sigma =0$ and $-x_3-\rho+\sigma =0$ are naturally interpreted as the endpoints of the two string segments moving along their axis towards each other at the speed of light until they reach the horizon, after which spacetime is described by the static NUT metric.
	}
	\label{planeWaves2}
\end{figure}

In Region III the metric takes a similar form with the replacements $u \longleftrightarrow v$, $q\rightarrow -q$ and $x_3 \rightarrow -x_3 $, and it therefore describes the upper half plane (or the northern hemisphere) with a semi-infinite string whose endpoint moves at the speed of light in the negative $x_3$ direction until it reaches the Taub-NUT horizon. The two semi-infinite strings reach the Taub-NUT horizon at the same time, after which spacetime is described by the static NUT metric (see figure \ref{planeWaves2}).

\begin{figure}[]
	\begin{center}
		\includegraphics[scale=1.6]{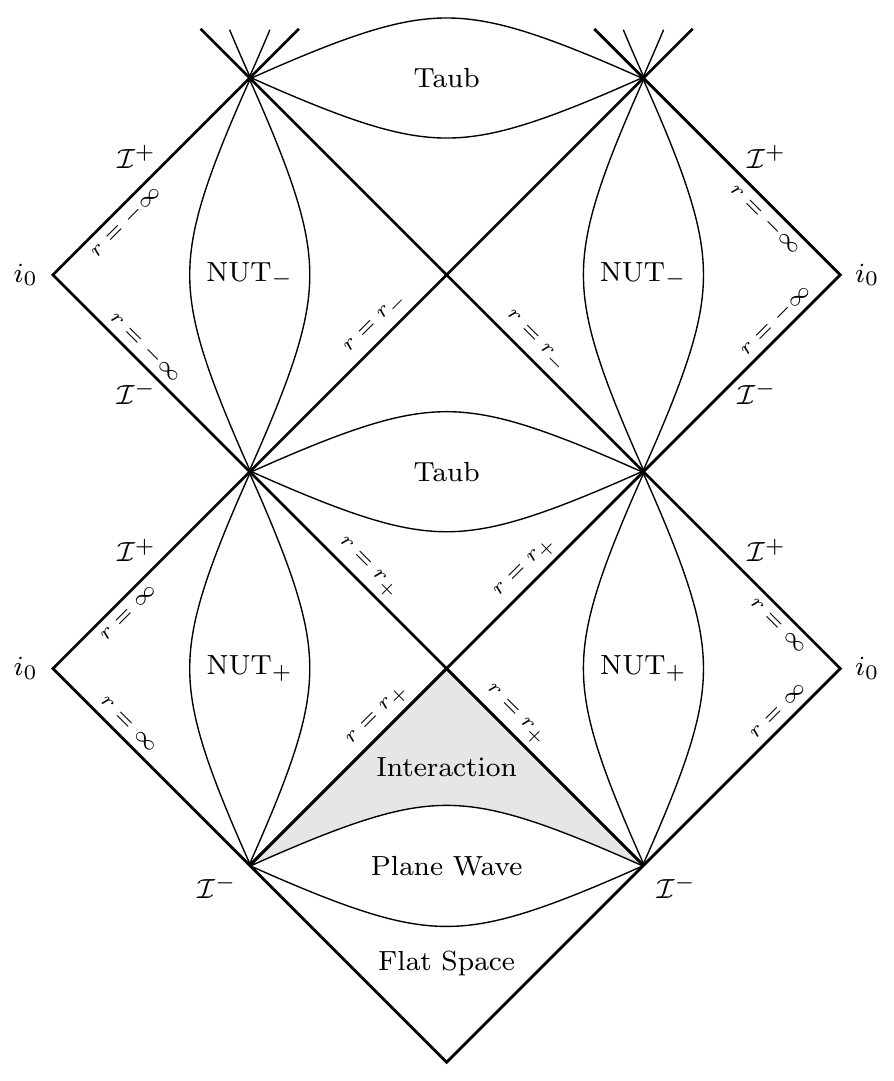}
	\end{center}
\vspace{-0.5cm}
	\caption{The plane waves solution in the $r$-$t$ plane. The lines drawn describe constant $r$ surfaces. The gray area describes the interaction region IV. The observer will propagate through the plane wave region II or III depending on its location on the $x_3$-axis ($x_3<0$ or $x_3>0$ respectively). Geodesics with $|x_3|>\sigma$ do not pass through the gray region, which shrinks to zero in this case; correspondingly, the plane wave region extends all the way to the horizon at $r=r_+$. In the special case $x_3=0$ the observer is equally separated from the two strings and therefore never propagates through the plane wave region, which shrinks to zero. The hypersurface that separates the interaction region from the flat space region in this case is $r=m$ (or equivalently $\rho = - \sigma$). Notice that some geodesic 
	in the plane wave region may
	end on the fold singularity, which does not appear in this 2D diagram.
	}
	\label{planeWaves3}
\end{figure}

 Finally, in figure \ref{planeWaves3} we draw a Penrose diagram of the solution in the $r$-$t$ plane.
The solution interpolates between regions of spacetime with different dual supertranslation charge. In the interaction region spacetime is locally isomorphic to the Taub-NUT metric and therefore it carries a dual supertranslation charge \eqref{TNcharge}. Before the instant of collision, on the other hand, spacetime is described by freely propagating plane waves that are neutral under dual supertranslations. The imaginary part of the boundary graviton $\Im C$ also changes as a result of the collision. In the interaction region it acquires a non-zero value, given by \eqref{ImCTN}, while it vanishes elsewhere.

The Taub-NUT solution is interpreted as a \emph{gravitomagnetic cosmic string}\footnote{By ``gravitomagnetic cosmic string" we mean a cosmic string with NUT charge. In the literature this is sometimes referred to as a ``spinning" cosmic string, but this term is also used to describe regular (Kerr) angular momentum. In order not to confuse between the NUT and Kerr charges we avoid using the term ``spinning."} (see \cite{Griffiths:2009dfa}, for example), which is the gravitational analogue of the Dirac string for magnetic monopoles. The solution described in this section is therefore naturally interpreted as the process of formation of a gravitomagnetic cosmic string (the reverse process of a snapping string that breaks into two parts).
A similar process that involves a snapping cosmic string was studied in \cite{Strominger:2016wns} as a solution that interpolates between vacua with different superrotation charges.
Let us stress that these two cosmic string solutions carry different charges (NUT/dual supertranslation versus Kerr angular momentum/superrotations).
In a similar way, the process of creation (or annihilation) of a gravitomagnetic cosmic string interpolates between vacua with different dual supertranslation charge.
This provides a physical interpretation of dual supertranslations.

 \section{Conclusions and Future Directions}\label{ConclusionSection}

 We have discussed several properties of the symmetry generated by the dual supertranslations charges 
 of~\cite{Godazgar:2018qpq,Godazgar:2018dvh}; we have shown in particular that the global dual supertranslation charge 
 is invariant under arbitrary globally defined, smooth deformation of the metric. The global dual supertranslation charge is 
 therefore a topological invariant and it classifies the topology of the two dimensional \emph{celestial space}, which is a 
 generalization of the celestial sphere. The trivial case is when spacetime has the topology of a sphere, in which case it is 
 \emph{globally} asymptotically flat and correspondingly the global dual supertranslation charge vanishes.
 When spacetime has a non-trivial topology, it is only \emph{locally} asymptotically flat.
 An example of a spacetime with a non-trivial topology is the Taub-NUT metric where, due to a cosmic 
 string defect the topology 
 of the celestial space is that of a branched covering of the sphere and correspondingly the dual 
 supertranslation charge is non-zero. In the same way, multi-NUT-string solutions will have the topology of a branched 
 covering of the sphere with multiple branch points located at the position of the strings.
 
 One may be tempted to draw a comparison with superrotations. Indeed, in both cases the violation of global asymptotic 
 flatness is due to singularities at isolated points on the celestial sphere.
 However, we would like to emphasize some key differences between the two symmetry operations.
The group of superrotation transformations is non-Abelian, which in turn implies that its generators do not commute 
with the Hamiltonian.
Moreover, superrotations produce a non-vanishing Bondi news at spatial infinity, indicating the emission of an infinite 
amount of gravitational radiation and therefore an infinite amount of energy.
Dual supertranslations, on the other hand, form an Abelian group, they commute with the Hamiltonian and therefore 
do not change the energy of the state and do not act on the Bondi news.
These key differences therefore suggest that the violation of global asymptotic flatness may be milder in the case of 
dual supertranslations.

The structure that we found in this work is analogous to the theory of magnetic monopoles in electrodynamics.
In the same way that magnetic charges classify the space of asymptotic gauge fields into distinct topological sectors, 
dual supertranslation charges classify the space of asymptotic metrics according to their topology.
 It is therefore natural to ask whether there exists a quantization condition for gravity similar to the Dirac quantization 
 condition in electrodynamics~\footnote{ See~\cite{comp2} for  a discussion on related topics.} . We hope to report on this issue soon.

Let us also remark that in this work we assumed that the Bondi mass is real. An imaginary Bondi mass will prevent the 
dual supertranslation charge from being conserved identically and is the gravitational analog of a magnetic source. 
It is an imaginary metric component that generates imaginary spacetime transformations. In contrast to QED, where a 
magnetic source generates imaginary gauge transformations that are perfectly understood, it seems harder to make 
sense of the gravitational analog. Nonetheless, it could still have interesting physical application, as in the description 
of thermal states. We leave these questions for future research.

 \section*{Acknowledgements}
MP thanks the Erwin Schr\"odinger Institute and the Galileo Galilei Institute for Theoretical Physics for its hospitality and 
INFN for partial support during the completion of this work. 
The research of UK and MP is supported in part by NSF through grant PHY-1620039.



\begin{thebibliography}{1}
\bibitem{Godazgar:2018qpq}
H.~Godazgar, M.~Godazgar and C.~N.~Pope,
``New Dual Gravitational Charges,''
arXiv:1812.01641 [hep-th].

\bibitem{Godazgar:2018dvh}
H.~Godazgar, M.~Godazgar and C.~N.~Pope,
``Tower of Subleading Dual BMS Charges,''
arXiv:1812.06935 [hep-th].

\bibitem{He:2014cra}
T.~He, P.~Mitra, A.~P.~Porfyriadis and A.~Strominger,
``New Symmetries of Massless Qed,''
JHEP {\bf 1410} (2014) 112
doi:10.1007/JHEP10(2014)112
[arXiv:1407.3789 [hep-th]].

\bibitem{Strominger:2013lka}
A.~Strominger,
``Asymptotic Symmetries of Yang-Mills Theory,''
JHEP {\bf 1407} (2014) 151
doi:10.1007/JHEP07(2014)151
[arXiv:1308.0589 [hep-th]].


\bibitem{Strominger:2015bla}
A.~Strominger,
``Magnetic Corrections to the Soft Photon Theorem,''
Phys.\ Rev.\ Lett.\ {\bf 116} (2016) no.3, 031602
doi:10.1103/PhysRevLett.116.031602
[arXiv:1509.00543 [hep-th]].




\bibitem{Strominger:2017zoo}
A.~Strominger,
``Lectures on the Infrared Structure of Gravity and Gauge Theory,''
arXiv:1703.05448 [hep-th].

\bibitem{shiu} Y.~Hamada, M.~S.~Seo and G.~Shiu,
  ``Large gauge transformations and little group for soft photons,''
  Phys.\ Rev.\ D {\bf 96}, no. 10, 105013 (2017)
  doi:10.1103/PhysRevD.96.105013
  [arXiv:1704.08773 [hep-th]];
  ``Electromagnetic Duality and the Electric Memory Effect,''
  JHEP {\bf 1802}, 046 (2018)
  doi:10.1007/JHEP02(2018)046
  [arXiv:1711.09968 [hep-th]].
 
\bibitem{Strominger:2013jfa}
A.~Strominger,
``On BMS Invariance of Gravitational Scattering,''
JHEP {\bf 1407} (2014) 152
doi:10.1007/JHEP07(2014)152
[arXiv:1312.2229 [hep-th]].

\bibitem{He:2014laa}
T.~He, V.~Lysov, P.~Mitra and A.~Strominger,
``BMS Supertranslations and Weinberg’s Soft Graviton Theorem,''
JHEP {\bf 1505} (2015) 151
doi:10.1007/JHEP05(2015)151
[arXiv:1401.7026 [hep-th]].





\bibitem{Henneaux:2018cst}
M.~Henneaux and C.~Troessaert,
``BMS Group at Spatial Infinity: the Hamiltonian (Adm) Approach,''
JHEP {\bf 1803} (2018) 147
doi:10.1007/JHEP03(2018)147
[arXiv:1801.03718 [gr-qc]].


\bibitem{Henneaux:2018gfi}
M.~Henneaux and C.~Troessaert,
``Asymptotic Symmetries of Electromagnetism at Spatial Infinity,''
JHEP {\bf 1805} (2018) 137
doi:10.1007/JHEP05(2018)137
[arXiv:1803.10194 [hep-th]].


\bibitem{Henneaux:2018hdj}
M.~Henneaux and C.~Troessaert,
``Hamiltonian Structure and Asymptotic Symmetries of the Einstein-Maxwell System at Spatial Infinity,''
JHEP {\bf 1807} (2018) 171
doi:10.1007/JHEP07(2018)171
[arXiv:1805.11288 [gr-qc]].







\bibitem{Barnich:2011mi}
G.~Barnich and C.~Troessaert,
``BMS Charge Algebra,''
JHEP {\bf 1112} (2011) 105
doi:10.1007/JHEP12(2011)105
[arXiv:1106.0213 [hep-th]].





\bibitem{Barnich:2016lyg}
G.~Barnich and C.~Troessaert,
``Finite BMS Transformations,''
JHEP {\bf 1603} (2016) 167
doi:10.1007/JHEP03(2016)167
[arXiv:1601.04090 [gr-qc]].


\bibitem{Choi:2017bna}
S.~Choi, U.~Kol and R.~Akhoury,
``Asymptotic Dynamics in Perturbative Quantum Gravity and BMS Supertranslations,''
JHEP {\bf 1801} (2018) 142
doi:10.1007/JHEP01(2018)142
[arXiv:1708.05717 [hep-th]].



\bibitem{Dirac:1931kp} 
P.~A.~M.~Dirac,
``Quantised singularities in the electromagnetic field,,''
Proc.\ Roy.\ Soc.\ Lond.\ A {\bf 133}, no. 821, 60 (1931).
doi:10.1098/rspa.1931.0130






\bibitem{Strominger:2014pwa}
A.~Strominger and A.~Zhiboedov,
``Gravitational Memory, BMS Supertranslations and Soft Theorems,''
JHEP {\bf 1601} (2016) 086
doi:10.1007/JHEP01(2016)086
[arXiv:1411.5745 [hep-th]].



\bibitem{Hawking:2016sgy}
S.~W.~Hawking, M.~J.~Perry and A.~Strominger,
``Superrotation Charge and Supertranslation Hair on Black Holes,''
JHEP {\bf 1705} (2017) 161
doi:10.1007/JHEP05(2017)161
[arXiv:1611.09175 [hep-th]].


\bibitem{comp} 
G.~Comp\`ere, A.~Fiorucci and R.~Ruzziconi,
  ``Superboost transitions, refraction memory and super-Lorentz charge algebra,''
  JHEP {\bf 1811}, 200 (2018)
  doi:10.1007/JHEP11(2018)200
  [arXiv:1810.00377 [hep-th]].

\bibitem{Pasterski:2015tva}
S.~Pasterski, A.~Strominger and A.~Zhiboedov,
``New Gravitational Memories,''
JHEP {\bf 1612} (2016) 053
doi:10.1007/JHEP12(2016)053
[arXiv:1502.06120 [hep-th]].










\bibitem{Ferrari:1988nu}
V.~Ferrari and J.~Ibanez,
``Type $D$ Solutions Describing the Collisions of Plane Fronted Gravitational Waves,''
Proc.\ Roy.\ Soc.\ Lond.\ A {\bf 417} (1988) 417.
doi:10.1098/rspa.1988.0068








\bibitem{Ferrari:1987cs}
V.~Ferrari and J.~Ibanez,
``On the Collision of Gravitational Plane Waves: a Class of Soliton Solutions,''
Gen.\ Rel.\ Grav.\ {\bf 19} (1987) 405.
doi:10.1007/BF00767280



\bibitem{Ferrari:1987yk}
V.~Ferrari and J.~Ibanez,
``A New Exact Solution for Colliding Gravitational Plane Waves,''
Gen.\ Rel.\ Grav.\ {\bf 19} (1987) 383.
doi:10.1007/BF00767279



\bibitem{Khan:1971vh}
K.~A.~Khan and R.~Penrose,
``Scattering of Two Impulsive Gravitational Plane Waves,''
Nature {\bf 229} (1971) 185.
doi:10.1038/229185a0






\bibitem{Newman:1961qr}
E.~Newman and R.~Penrose,
``An Approach to Gravitational Radiation by a Method of Spin Coefficients,''
J.\ Math.\ Phys.\ {\bf 3} (1962) 566.
doi:10.1063/1.1724257


\bibitem{Newman:1965ik}
E.~T.~Newman and R.~Penrose,
``10 Exact Gravitationally-Conserved Quantities,''
Phys.\ Rev.\ Lett.\ {\bf 15} (1965) 231.
doi:10.1103/PhysRevLett.15.231


\bibitem{Newman:1966ub}
E.~T.~Newman and R.~Penrose,
``Note on the Bondi-Metzner-Sachs Group,''
J.\ Math.\ Phys.\ {\bf 7} (1966) 863.
doi:10.1063/1.1931221


\bibitem{Newman:1968uj}
E.~T.~Newman and R.~Penrose,
``New Conservation Laws for Zero Rest-Mass Fields in Asymptotically Flat Space-Time,''
Proc.\ Roy.\ Soc.\ Lond.\ A {\bf 305} (1968) 175.
doi:10.1098/rspa.1968.0112




\bibitem{Talbot:1969bpa}
C.~J.~Talbot,
``Newman-Penrose Approach to Twisting Degenerate Metrics,''
Commun.\ Math.\ Phys.\ {\bf 13} (1969) no.1, 45.
doi:10.1007/BF01645269


\bibitem{Penrose:1986ca}
R.~Penrose and W.~Rindler,
``Spinors and Space-Time. Vol. 2: Spinor and Twistor Methods in Space-Time Geometry,''
doi:10.1017/CBO9780511524486



\bibitem{Penrose:1987uia}
R.~Penrose and W.~Rindler,
``Spinors and Space-Time,''
doi:10.1017/CBO9780511564048




\bibitem{Adamo:2009vu}
T.~M.~Adamo, C.~N.~Kozameh and E.~T.~Newman,
``Null Geodesic Congruences, Asymptotically Flat Space-Times and Their Physical Interpretation,''
Living Rev.\ Rel.\ {\bf 12} (2009) 6
[Living Rev.\ Rel.\ {\bf 15} (2012) 1]
doi:10.12942/lrr-2009-6
[arXiv:0906.2155 [gr-qc]].




\bibitem{Adamo:2014baa}
T.~Adamo and E.~T.~Newman,
``The Kerr-Newman Metric: a Review,''
Scholarpedia {\bf 9} (2014) 31791
doi:10.4249/scholarpedia.31791
[arXiv:1410.6626 [gr-qc]].














\bibitem{Matzner:1984pe}
R.~A.~Matzner and F.~J.~Tipler,
``Metaphysics of Colliding Selfgravitating Plane Waves,''
Phys.\ Rev.\ D {\bf 29} (1984) 1575.
doi:10.1103/PhysRevD.29.1575





\bibitem{Griffiths:1991zp} 
J.~B.~Griffiths,
``Colliding plane waves in general relativity,''
Oxford, UK: Clarendon (1991) 232 p. (Oxford mathematical monographs)









\bibitem{Griffiths:2009dfa} 
J.~B.~Griffiths and J.~Podolsky,
``Exact Space-Times in Einstein's General Relativity,''
doi:10.1017/CBO9780511635397






\bibitem{Strominger:2016wns}
A.~Strominger and A.~Zhiboedov,
``Superrotations and Black Hole Pair Creation,''
Class.\ Quant.\ Grav.\ {\bf 34} (2017) no.6, 064002
doi:10.1088/1361-6382/aa5b5f
[arXiv:1610.00639 [hep-th]].

\bibitem{comp2}
C.~W.~Bunster, S.~Cnockaert, M.~Henneaux and R.~Portugues,
  ``Monopoles for gravitation and for higher spin fields,''
  Phys.\ Rev.\ D {\bf 73}, 105014 (2006)
  doi:10.1103/PhysRevD.73.105014
  [hep-th/0601222].



  
 \end{thebibliography}
\end{document}